\documentclass[12pt,epsfig]{article}
\usepackage{graphicx}
\def\beq{\begin{equation}}
\def\eeq{\end{equation}}
\def\bea{\begin{eqnarray}}
\def\eea{\end{eqnarray}}

\begin{document}
% \begin{flushright} hep-th/9701097
% \end{flushright}
% \rightline{TIFR/TH/97-01}
% \rightline{January 1997}

\begin{center}
{\Large \bf Hartman-Fletcher effect for array of complex barriers
  }

\vspace{1.3cm}

{\sf  Ananya Ghatak \footnote{e-mail address: \ \ gananya04@gmail.com}, Mohammad Hasan \footnote{e-mail address: \ \ mohammadhasan786@gmail.com}
and Bhabani Prasad Mandal \footnote{e-mail address:
\ \ bhabani.mandal@gmail.com, \ \ bhabani@bhu.ac.in  }}

\bigskip

{\em $^{1,3}$Department of Physics,
Banaras Hindu University,
Varanasi-221005, INDIA. \\
$^2$ISRO Satellite Centre (ISAC),
Bangalore-560017, INDIA \\ }

\bigskip
\bigskip

\noindent {\bf Abstract}

\end{center}
We calculate the time taken by a wave packet to tunnel through a series of complex
barrier potentials using stationary phase method to show its saturation (Hartman-Fletcher effect) with number of 
barriers in various situations. We numerically study the effect of the
coupling between the elastic and inelastic channels,
width of the individual barrier, separation between the consecutive barriers on the
saturation of tunneling time. Nature of HF effect
has further been investigated for more realistic barriers with random inelasticity and also
for emissive inelastic channels.

\medskip
\vspace{1in}
\newpage

\section{Introduction}
The propagation of an evanescent wave through a potential barrier has long been studied \cite{htt1,
htt2,htt3,htt4}. The question of tunneling time i.e. the
time spent by a wave packet with mean incident energy smaller than barrier height in the classically forbidden region and the analogous  
situation of evanescent waves in optics have in recent years attracted considerable attention. One of the most interesting aspect
of these studies is the saturation of tunneling time with respect to the width of the barrier and is referred as Hartman-Fletcher (HF) effect. Hartman studied \cite{hart} the tunneling time by constructing a metal-insulator-metal 
sandwich by the method of stationary phase to demonstrate the experimental agreement of the 
in dependency of tunneling time on the width of the barrier. Fletcher independently showed \cite{flet} 
the saturation of time delay by considering tunneling of evanescent wave through a thick barrier. 
This exciting result triggers series of famous experiments with incident wave both in microwave \cite{exp1}-\cite{exp4} and optical range \cite{exp5}-\cite{exp6} to obtain the saturation of tunneling time
by considering single, double, and multiple real barrier potentials. Recently Longhi et. al
\cite{exp6} have measured tunneling time for a double barrier optical
grating and found that the tunneling time is paradoxically short and independent of barrier width and 
separation between the barriers. Similar conclusion was obtained by Dutta-Roy et. al \cite{dr} by 
considering single barrier associated with inelasticity.

In the present work we study the HF effect by considering an array of non-Hermitian barrier
potentials. The motivation arises from the huge applicability of non-Hermitian system
\cite{ben4}-\cite{benr} in various 
branches of physics over the past one and half decades. In particular non-Hermitian theories are considered to be the topic of frontier research work on transport \cite{ent}-\cite{cal} and
scattering \cite{dp}-\cite{hsn} phenomena for matter as well as electromagnetic waves.
Scattering from complex potential has very rich features and leads to many technological developments \cite{cpa011, cpa00, cpa01}. Even though the HF effect have been extensively studied both theoretically \cite{nimz}-\cite{win} and experimentally \cite{exp1}-\cite{exp6} for real periodic potentials,
it has not been discussed for complex potentials which are associated with both elastic and 
inelastic channels, except the work in \cite{dr} where the tunneling time for a single complex barrier has been 
calculated and HF effect is discussed for the weak absorption. 
For the strong absorption the HF effect for a single complex 
barrier is shown to disappear. This result is consistent with the experimental findings in Ref.\cite{exp1}.
Being a quantum mechanical process the characteristics of tunneling do
not guarantee to be the same when wave transmits through multiple number of barriers which are separated from one another. Therefore
it is worth investigating the various characteristics of tunneling time and the presence of 
HF effect when waves scatter through an array of complex
barriers. We consider an array of non-Hermitian square barrier potentials with a fixed height and periodicity which is actually the mathematical structure of the famous Cronning-Penny model that depicts the lattice structures in crystal.
In this work we calculate the time taken by a wave packet to tunnel through such an array of complex 
barrier potentials to show the various characteristics of HF effect. 
We put some light on the
the behavior of tunneling time and discuss the
long debated HF effect i.e. the saturation of tunneling time for matter waves scatter through such arrays
of non-Hermitian potentials.
We calculate the tunneling time by using the method of stationary phase for an array consist of any 
number of complex barrier potentials and discuss the saturation of tunneling time with the number of the 
complex barriers by varying different parameters in the system. The saturation crucially depends
on the coupling $V_c$ between elastic and inelastic channels. Saturation is achieved only for small $V_c$.
This implies system shows HF effect only for less absorption which is consistent with the results in \cite{exp1, dr}.
For fixed
$V_c$ HF effect depends on the width of the individual barrier
and separation between the consecutive barriers. Unlike the array of real barriers here the
saturation is obtained only for certain range of width of individual barrier.
 One of the rich features of tunneling time for real barriers is the occurrence of resonances \cite{rns1}
for specific values of the separation between adjacent barriers. We observe some of resonances disappear with increase
of absorptivity in the system of array of complex barriers.
We show that the resonances in tunneling time for array of complex barriers is regulated with increasing $V_c$.
We further demonstrate the existence of HF effect even for more realistic situations where inelasticity
is randomly chosen for individual barriers.
In case of emissive inelastic channel we show that the HF effect occurs only at certain small values of incident energy.

Now we present the plan of this paper. We review the HF effect for single barrier
in Sec.2 to outline the methodology for later sections. In Sec.3 tunneling
time is calculated for an array of complex barriers. Various results regarding the HF effect
for arrays of complex barriers are discussed in Sec.4. Sec.5 is kept for conclusion and remarks on 
further prospectives.

\section{HF effect for single barrier}

\begin{figure}
\centering 

\includegraphics[scale=0.82]{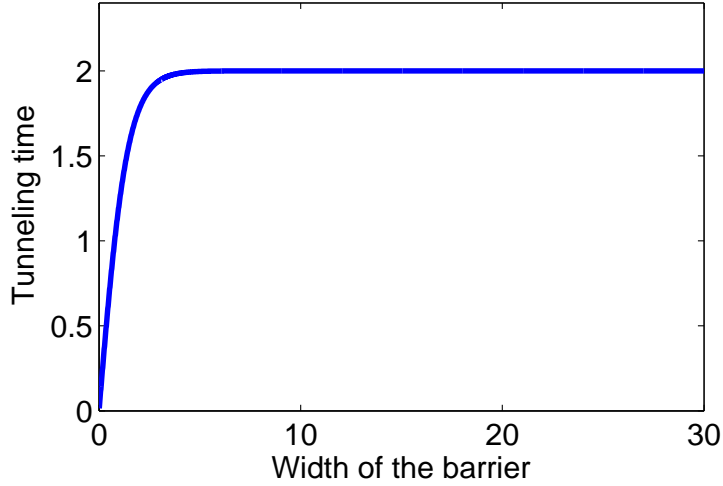} 

\caption {\it Tunneling time with the width of the barrier shows Hartman-Fletcher effect for a real barrier 
with $ V_r=1$, $E=0.5$ and $\hbar=1=2m$.} 
\label{8.1}
\end{figure}

In this section we review the method of calculating the tunneling time of a wave packet through a single real 
barrier by stationary 
phase method \cite{bnkb}. In this method the tunneling time ($\tau$) is defined as the time taken by the peak 
of the incident wave packet to traverse the classically forbidden region and emerges as transmitted wave packet.
To calculate $\tau$  we consider the evolution of an incident localized wave packet which is described by,
\beq\int G_{k_0}(k)exp(i k x -i E t/ \hbar)dk \ .
\eeq
$G_{k_0}(k)$ is the normalized Gaussian function of wave number $k$ peaked about the mean momentum $\hbar k_0$. 
Due to presence of a barrier an incident wave packet would emerge after transmission as 
\beq
\int G_{k_0}\mid a_T \mid exp(i k x -i E t/ \hbar +i \Phi (k))dk
\eeq
where $a_T$ is the transmitted amplitude $a_T=\mid a_T(k) \mid e^{i \Phi(k)}$. The time $\tau$ at which the peak 
of the  wave packet
emerges from the barrier ($V=V_r$ for $0\leq x\leq b$ and $V=0$ elsewhere) is the given by stationary phase method as,
\beq
\frac{d}{dk}[i k b -i E \tau/ \hbar +i \Phi (k)]=0 \ .
\eeq
This implies
\beq
\tau =\hbar \frac{d\Phi (E)}{dE}+\frac{b}{\hbar k /m} \ .
\label{tt}
\eeq
For a single real barrier the transmission amplitude is given by,
\beq
a_T(k)=\frac{2 k q e^{-i b k}} {2 k q \cosh( b q) - i (2 E - V_r) \sinh( b q)} \ \ \mbox{with} \ \ k=\sqrt{2mE}, \ q=\sqrt{2m(V_r-E)}/\hbar
\label{ta}
\eeq 
Thus $\tau$ is calculated in stationary phase method \cite{bnkb} as,
\beq
\tau =\hbar\frac{d}{dE}\tan^{-1}\left [\frac{k^2-q^2}{2qk}\tanh{qb}\right ], 
\label{tt2}
\eeq
From the above equation we observe that $\tau \rightarrow 0$ as $b\rightarrow 0$ as expected,
however when $b\rightarrow \infty$, $\tau =\frac{2m}{\hbar k q}$, i.e. tunneling time is independent of the width
of the barrier $b$ for sufficiently opaque barrier. This paradoxical result (HF Effect) is demonstrated in 
Fig. \ref{8.1}.

The Fig. \ref{8.1} shows the saturation of tunneling time when width of 
the barrier is larger than a certain value and incident energy is less enough than the height of the barrier. Recently the tunneling time
has also been calculated for single complex barrier potential to show the HF effect \cite{dr} for weak absorption. HF effect disappears
when the imaginary part of the barrier has large value.
In the later sections we discuss the tunneling time and HF effect for array of complex barriers.

\section{Tunneling time for array of complex barriers}
In this section we intend to calculate tunneling time for an array of complex barriers. For this purpose we start 
with a single complex barrier with a non-Hermitian potential,
\bea
V(x)&=& V_r +i V_i \ \ \ \mbox{for} \ \ a_1\leq x\leq a_2 \ \ \nonumber \\
&=& 0 \ \ \ \ \ \ \ \ \mbox{elsewhere} \ ,
\label{pot2}
\eea
where $V_r$ and $ V_i$ are real and $a_2-a_1=b$ is the width of the barrier.
Following the idea of \cite{dr, twochan} this complex barrier can physically be realized by two channel formalism, where an absorptive (or emissive) inelastic channel is evanescently coupled with the elastic channel via a coupling potential $V_c$.  
The real part of the potential is corresponding to an elastic channel described by the
Schroedinger equation (with $\hbar=1=2m$), 
\beq
\left [-\frac{d^2}{dx^2}+V_r\right ]\psi(x)+V_c\phi(x)=E\psi(x) ,
\label{el}
\eeq
Whereas the imaginary part of the potential is associated with an inelastic channel and is
described by the Schroedinger equation as,
\beq
\left [-\frac{d^2}{dx^2}+V_i\right ]\phi(x)+V_c\psi(x)=(E-\Delta )\psi(x)
\label{inel}
\eeq
$\Delta$ is the energy absorbed by the system from incident wave due to the presence of the 
imaginary part of the potential. For positive $\Delta$, the inelastic channel is absorptive whereas
it is emissive for negative $\Delta$.
The scattering wave functions in the elastic and inelastic 
 channels (as in Ref. \cite{dr}) can be obtained by solving Eq. (\ref{el}) and (\ref{inel}) successively,
\bea
\psi(x)&=&(Be^{\alpha x}+C e^{-\alpha x})\sin (\theta /2) +(Fe^{\beta x}+G e^{-\beta x})\cos (\theta /2) \nonumber \\
\phi(x)&=&(Be^{\alpha x}+C e^{-\alpha x})\cos (\theta /2) -(Fe^{\beta x}+G e^{-\beta x})\sin (\theta /2)
\label{wv}
\eea
where, \beq
\alpha ^2=V_r-E+\frac{V_r-V_i-\Delta}{2}(\sec\theta -1) \ ; \
\beta ^2=V_r-E-\frac{V_r-V_i-\Delta}{2}(\sec\theta -1) \ ; \eeq
\beq \mbox{and} \ \ \theta=\tan ^{-1}\left(\frac{2 V_c}{V_r-V_i-\Delta}\right ) 
\label{teta}
\eeq
The asymptotic forms of these scattering wave functions in the elastic and inelastic 
 channels are given as follows,
\bea
&\mbox{at}& \ x<a_1, \ \ \psi^-(x)=Ae^{ikx}+De^{-ikx} \ \ ; \ \ \phi^-(x)=R_{inel}e^{-ik'x} 
\ , \nonumber \\
&\mbox{at}& \ x>a_2, \ \ \psi^+(x)=Pe^{ikx}+Qe^{-ikx} \ \ ; \ \ \phi^+(x)=T_{inel}e^{ik'x}
\label{asym}
\eea
where $k=\sqrt{E},\ k'=\sqrt{E-\Delta}$. 
As the wave functions of both the channels in Eq. 
(\ref{wv}), and their derivative satisfy the condition of continuity at $x>a_1$ and $x<a_1$,
one get total 8 equation for the two individual channels which are,
\beq
A e^{i a_1 k}+D e^{-i a_1 k}= \left(B e^{\alpha a_1}+C e^{-\alpha a_1}\right)\sin (\theta /2)+ \left(F e^{\beta a_1 }+G e^{- \beta a_1}\right)\cos (\theta /2)
\label{e1}
\eeq
\beq
i k(A e^{i a_1 k}-D e^{-i a_1 k})= \alpha \left(B e^{\alpha a_1}-C e^{-\alpha a_1}\right)\sin (\theta /2)+ \beta \left(F e^{\beta a_1 }-G e^{- \beta a_1}\right)\cos (\theta /2)
\eeq
\beq
P e^{i a_2 k}+Q e^{-i a_2 k}= \left(B e^{\alpha a_2}+C e^{-\alpha a_2}\right)\sin (\theta /2)+ \left(F e^{\beta a_2}+G e^{- \beta a_2}\right)\cos (\theta /2)
\eeq
\beq
i k(P e^{i a_2 k}+Q e^{-i a_2 k})= \alpha \left(B e^{\alpha a_2}-C e^{-\alpha a_2}\right)\sin (\theta /2)+ \beta \left(F e^{\beta a_2}-G e^{- \beta a_2}\right)\cos (\theta /2)
\eeq

\vspace{.05in}

\beq
R_{inel} e^{-i a_1 k'}= \left(B e^{\alpha a_1}+C e^{-\alpha a_1}\right)\cos (\theta /2)- \left(F e^{\beta a_1}+G e^{- \beta a_1}\right)\sin (\theta /2)
\eeq
\beq
-i k' R_{inel} e^{-i a_1 k'}= \alpha \left(B e^{\alpha a_1}-C e^{-\alpha a_1}\right)\cos (\theta /2)-\beta \left(F e^{\beta a_1}-G e^{- \beta a_1}\right)\sin (\theta /2)
\eeq
\beq
T_{inel} e^{i a_2 k'}= \left(B e^{\alpha a_2}+C e^{-\alpha a_2}\right)\cos (\theta /2)- \left(F e^{\beta a_2}+G e^{- \beta a_2}\right)\sin (\theta /2)
\eeq
\beq
ik'T_{inel} e^{i a_2 k'}= \alpha \left(B e^{\alpha a_2}-C e^{-\alpha a_2}\right)\cos (\theta /2)- \beta \left(F e^{\beta a_2}-G e^{- \beta a_2}\right)\sin (\theta /2)
\label{e8}
\eeq

During a bidirectional scattering in the elastic channel the asymptotic amplitudes in
the right hand side (i.e. $P, Q$) and those for the left hand side (i.e. $A, D$) are related to each others 
via the transfer matrix as,
\beq
\left(\begin{array}{clcr}
P   \\
Q  \\ \end{array}\right) = \left(\begin{array}{clcr}
M_{11} \ & M_{12}   \\
M_{21} \ & M_{22}   \\ \end{array}\right) \left(\begin{array}{clcr}
A   \\
D  \\
\end{array}\right) 
\label{pqab}
\eeq
From Eqs. (\ref{e1})-(\ref{e8}) and by using Eq. (\ref{pqab}) we calculate all the components of 
the M-matrix as,
\bea
M_{11}&=&\frac{e^{i(a_1-a_2)k}}{4k\alpha \beta }\left [ \alpha (ik^2z+2ky\beta -iz\beta ^2)
\cos^2(\theta /2)+\beta  (ik^2v+2kw\alpha  -iv\alpha ^2)\sin^2(\theta /2) \right. \nonumber \\
&-& \left. \frac{i(kk'z\alpha -(kk'v+i(k-k')(w-y)\alpha +v\alpha ^2)\beta +z\alpha \beta ^2)^2
\sin^2\theta }{4\beta  (k'^2v+2ik'w\alpha -v\alpha ^2)\cos^2(\theta /2)+4\alpha (k'^2z+2ik'y
\beta -z\beta ^2)\sin^2(\theta /2)}\right ]
\label{m11}
\eea
\bea
M_{12}&=&\frac{e^{-i(a_1+a_2)k}}{4k\alpha \beta }\left [ -iz\alpha (k^2+\beta ^2)
\cos^2(\theta /2)-iv\beta  (k^2+\alpha ^2)\sin^2(\theta /2) \right. \nonumber \\
&+& \left. \frac{i\left(k^2\left \{k'(z\alpha -v\beta )-i(w-y)\alpha \beta \right\}^2-
\alpha^2 \beta ^2\left\{v\alpha -z\beta -ik'(w-y)\right\}^2\right)
\sin^2\theta }
{4\beta  (k'^2v+2ik'w\alpha -v\alpha ^2)\cos^2(\theta /2)+4\alpha (k'^2z+2ik'y
\beta -z\beta ^2)\sin^2(\theta /2)}\right ] \ \ \ \ \ \ \ \ 
\label{m12}
\eea
\bea
M_{21}&=&\frac{e^{i(a_1+a_2)k}}{4k\alpha \beta }\left [ iz\alpha (k^2+\beta ^2)
\cos^2(\theta /2)+iv\beta  (k^2+\alpha ^2)\sin^2(\theta /2) \right. \nonumber \\
&-& \left. \frac{-i\left(k^2\left \{k'(z\alpha -v\beta )+i(w-y)\alpha \beta \right\}^2-
\alpha^2 \beta ^2\left\{v\alpha -z\beta -ik'(w-y)\right\}^2\right)
\sin^2\theta }
{4\beta  (k'^2v+2ik'w\alpha -v\alpha ^2)\cos^2(\theta /2)+4\alpha (k'^2z+2ik'y
\beta -z\beta ^2)\sin^2(\theta /2)}\right ] \ \ \ \ \ \ \ \ 
\label{m21}
\eea
\bea
M_{22}&=&\frac{e^{i(a_2-a_1)k}}{4k\alpha \beta }\left [ \alpha (-ik^2z+2ky\beta +iz\beta ^2)
\cos^2(\theta /2)+\beta  (-ik^2v+2kw\alpha  +iv\alpha ^2)\sin^2(\theta /2) \right. \nonumber \\
&+& \left. \frac{i(kk'z\alpha +(-kk'v-i(k+k')(w-y)\alpha +v\alpha ^2)\beta -z\alpha \beta ^2)^2
\sin^2\theta }{4\beta  (k'^2v+2ik'w\alpha -v\alpha ^2)\cos^2(\theta /2)+4\alpha (k'^2z+2ik'y
\beta -z\beta ^2)\sin^2(\theta /2)}\right ]
\label{m22}
\eea
where the notations $v = 2\sinh(\alpha b); w = 2\cosh(\alpha b); y = 
 2\cosh(\beta  b); z = 2\sinh(\beta  b)$ have been used.
Thus we find all the left
and right handed scattering amplitudes as, 
\beq
r_l=\frac{{M_{21}}}{{M_{22}}} \ \ ; \ \ r_r=\frac{{M_{12}}}{{M_{22}}};
\ \ t_l=\frac{det[M]}{{M_{22}}} \ \ ; \ \ t_r=\frac{1}{{M_{22}}} ; \ \ \mbox{with} \ \ det[M]=1 \ ;
\label{sc}
\eeq
and the transmission and reflection coefficients are obtained from these amplitudes as,
\beq
T_l=\mid t_l\mid^2=T_r \ ; \ R_l=\mid r_l\mid^2 \ ; \ R_l=\mid r_l\mid^2 \ .
\eeq
The phase difference between incident and transmitted waves is calculated from the 
transmitted amplitude in Eq. \ref{sc}. With this phase difference the tunneling time
for a single complex barrier is calculated by using Eq. \ref{tt}. As a check we chose $V_c=0$ for which we should 
recover the tunneling time for single real barrier. At this limiting case $M_{22}$ is reduced (as $\theta =0$ in Eq. (\ref {teta})) to,
\beq
M_{22}=\frac{e^{i b \sqrt{E}} \left[2 \sqrt{E} \sqrt{V_r-E } \cosh( b \sqrt{V_r-E}) - 
   i (2 E - V_r) \sinh( b \sqrt{V_r-E})\right]}{2 \sqrt{E} \sqrt{V_r-E}}
\eeq
due to which the transmission amplitudes in Eq. (\ref{sc}) is reduced to Eq. (\ref{ta}) and so the tunneling
time for a single real barrier can be re-obtained and the two channel 
study for the complex barrier is being verified. The tunneling time 
for a single complex barrier shows HF 
effect in the case of weak coupling (i.e. small $V_c$) which is already discussed in \cite{dr}.

We are now at the position to calculate the tunneling time for an array 
\cite{gr1,gr2,gr3} of complex barriers. 
For that we consider an array consisting $n$ complex barriers, each of width $b$ and 
consecutively gaped by a length $L$. Each barrier potential in the array
has been expressed in the same way as written in Eq. \ref{pot2} but now the span of
the $(n+1)^{th}$ potential will be decided by,
\beq
a_1=n(b+L) \ ; \ a_2=b+a_1 \ ; \ n=0,1,2,3... . 
\label{a1a2}
\eeq
We derive the elements of individual transfer matrices for the $n_{th}$ barrier potential
using Eqs. (\ref{m11}-\ref{m22}) and by replacing the corresponding $a_1, a_2$ as per in Eq. (\ref{a1a2}).  
We denote the  M-matrix for the first barrier (i.e. $n=0$, span is between 0 to b) as $M_0$
for the second barrier (i.e. $n=1$, span is between b+L to 2b+L) as $M_1$ and so on.
In this way the total transfer matrix for an array of $n$ barriers is written as,
\beq
M^{tot}=M_{n-1}.....M_3.M_2.M_1.M_0= \Pi^{j=0}_{n-1}M_j
\label{mm}
\eeq
Therefore the transmission amplitude for the array of $n$ barriers is 
$t^{tot}=\frac{1}{{M^{tot}_{22}}}$. Now this
transmission amplitude can be written also as $t^{tot}=\mid t^{tot} \mid e^{i \delta (k)}$ where $\delta 
(k)$ is the phase difference between incident and transmitted waves for the array of barriers. 
Now the total span of the interacting potential is the sum of all the individual barriers of width 
$b$ and their consecutive separations of length $L$. We can think the array of barriers equivalently as
a single potential distribution \cite{gr2} of width $nb+(n-1)L$.
Then using Eq. (\ref{tt}) we obtain the tunneling time for the
array of barriers as, 
\beq
\tau =\hbar\frac{d\delta (E)}{dE}+\frac{nb+(n-1)L}{\hbar k /m}
\label{tt3}
\eeq
As the matrix multiplication in Eq. (\ref{mm}) is too lengthy to handle 
analytically we numerically extract the behavior of tunneling time to see the
presence of HF effect in various situations. In the following sections we have demonstrated the results related to
HF effect in tunneling time for array of complex barriers.

\section{Results and Discussions}
In this section we consider the expression of $M^{tot}$ in Eq. (\ref{mm}) to find the effect of various parameters on the tunneling time. We consider both absorptive as well as emissive ($\Delta$ is negative) inelastic channels
in our discussion to observe HF effect with respect to number of barriers
by varying the coupling, width of the individual barrier and
separation between the adjacent barriers. 
We further discuss the tunneling time resonances in the array of complex barriers. For the sake of more realistic 
system we consider the coupling $V_c$ in a random manner for different barriers in one array and in another one we chose random excitation for different barriers. We are also able to capture the HF effect even in such 
realistic systems.

\subsection{Effect of coupling}
\begin{figure}
\centering 

\includegraphics[scale=0.48]{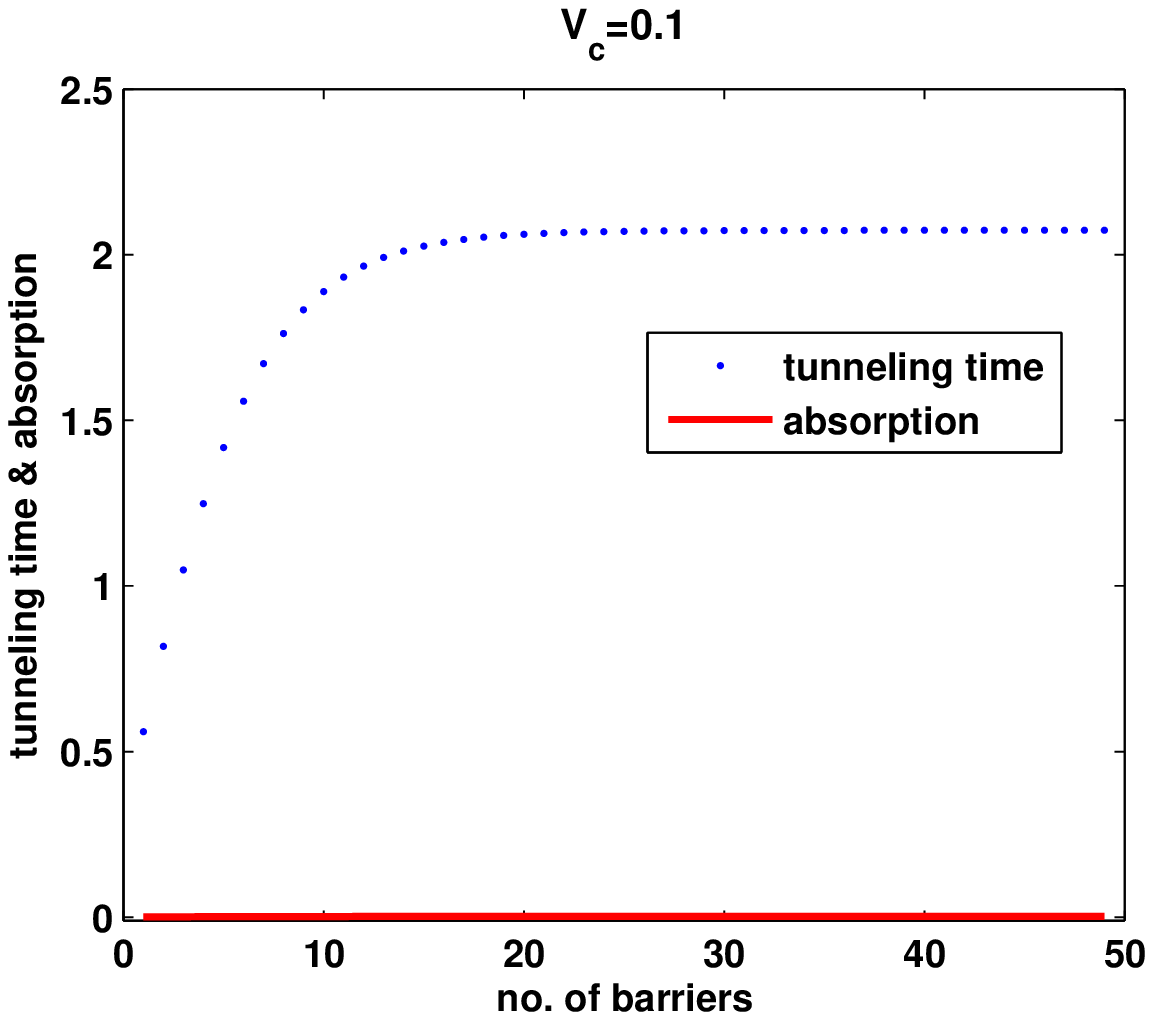} (a) \ \ \ \ \ \ \includegraphics[scale=.48]{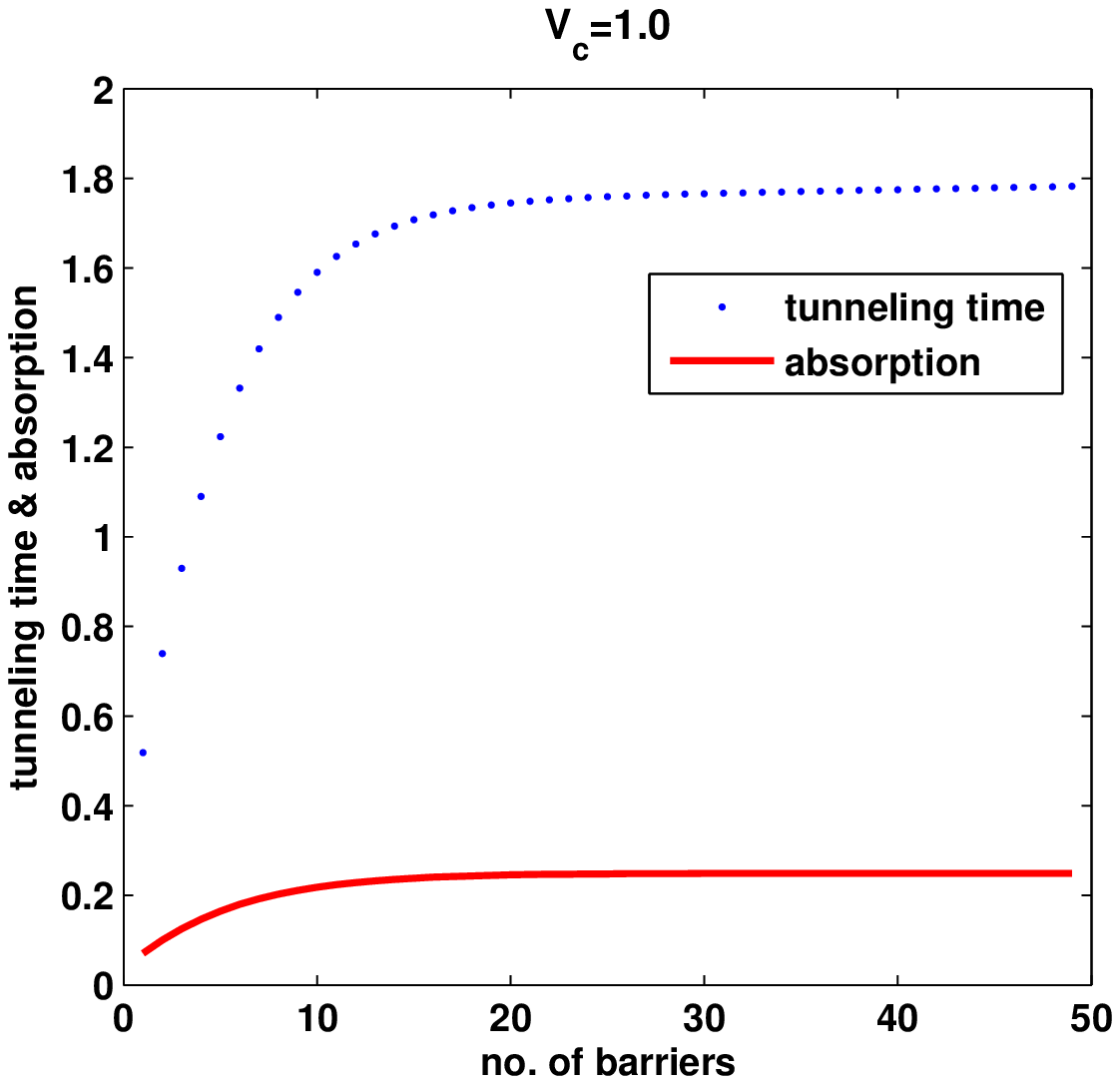} (b)

\includegraphics[scale=0.48]{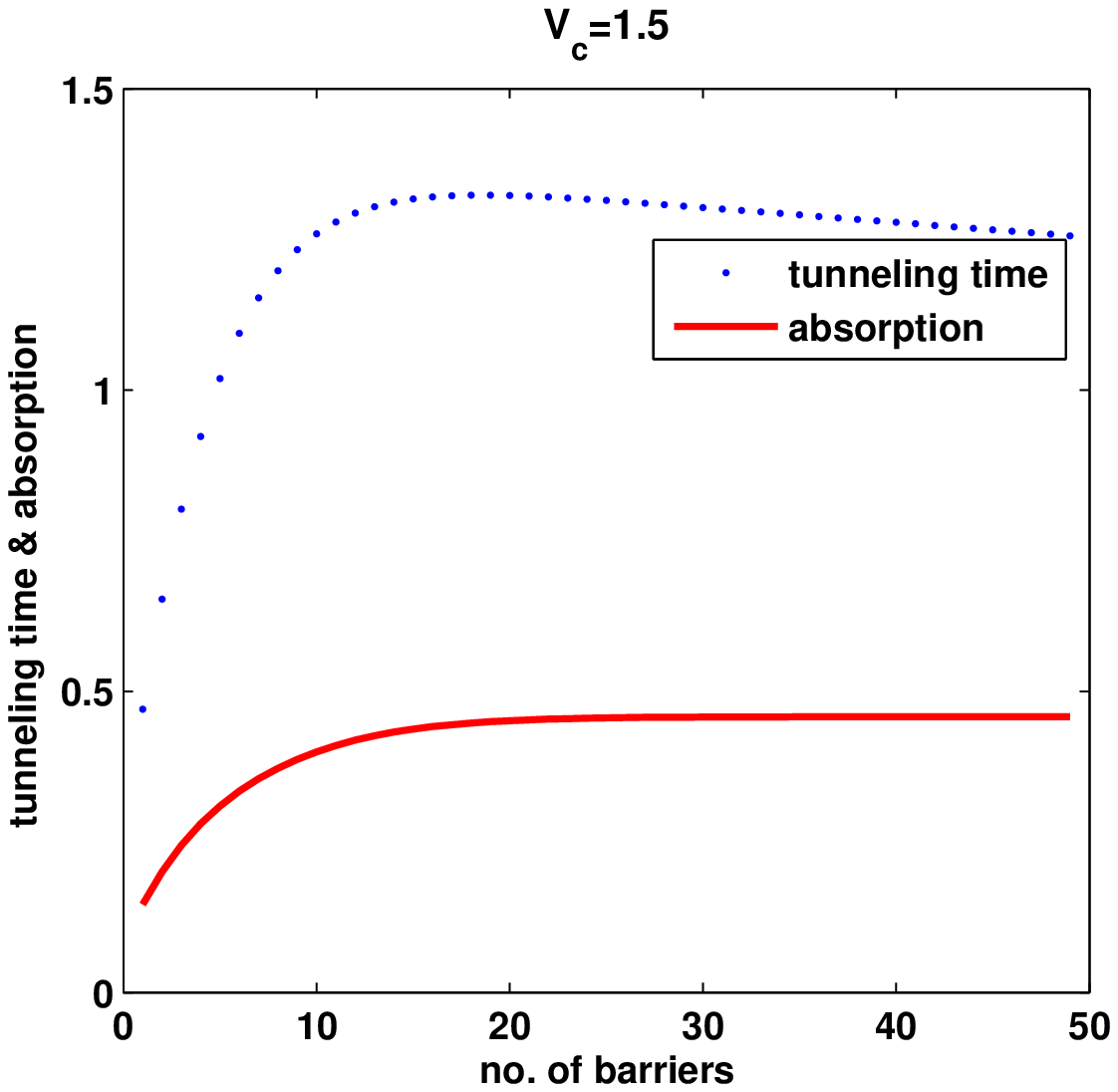} (c) \ \ \ \ \ \ \includegraphics[scale=.48]{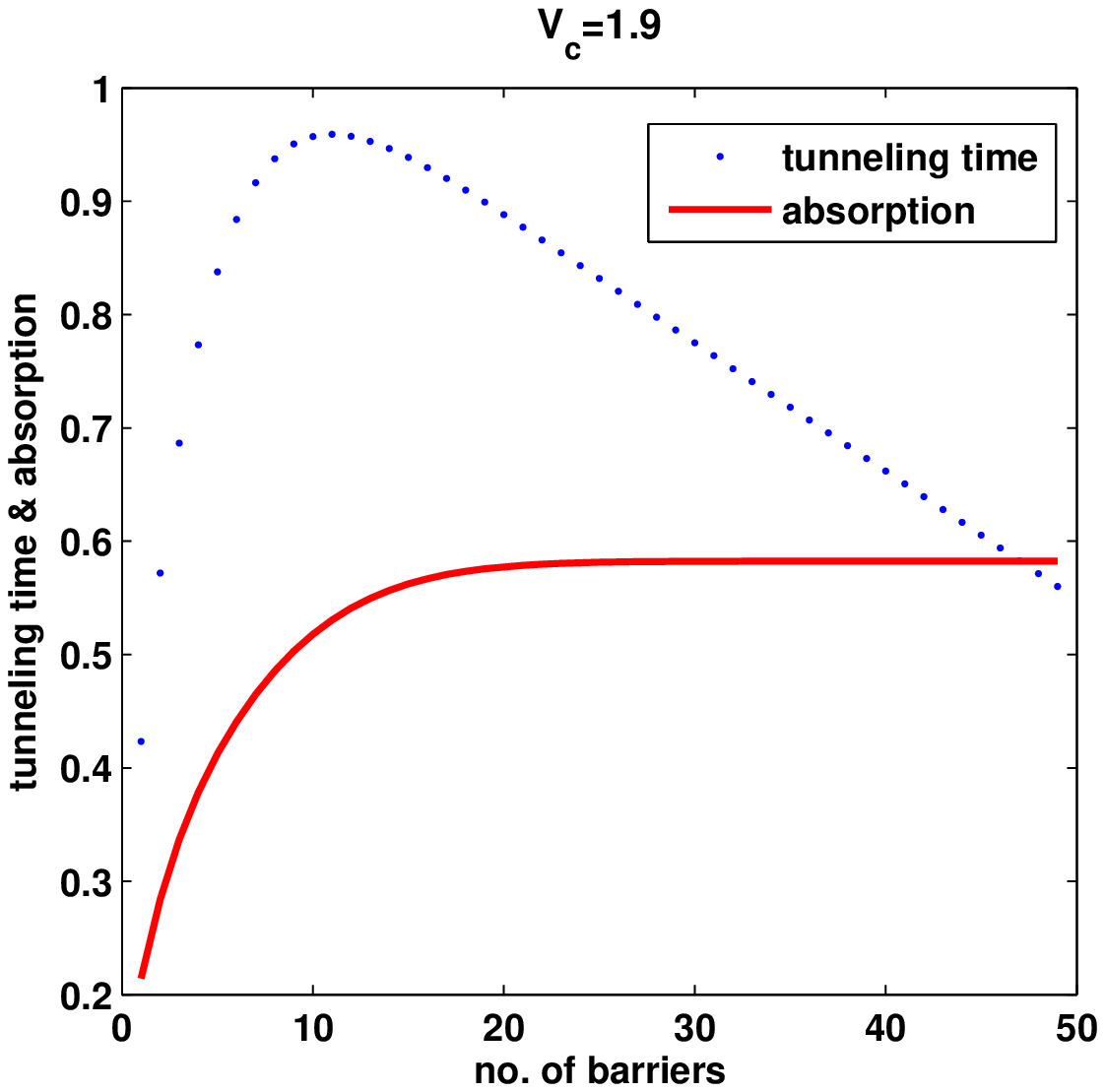} (d)
%\vspace{-1.0in}
%\hspace{-0.4in}\includegraphics[scale=0.82]{vc1p5.eps} \hspace{-1.8in} \includegraphics[scale=.8]{vc1p9.eps}

\caption {\it Tunneling times and absorptivity are shown with the number of barriers for four
different arrays with different values of $V_c$ with $E=0.5, V_r=1, V_i=1,\Delta=0.02, b=0.2, L=0.01$. In 
Fig.1(a)-(c) absorption saturates almost at low value and we observe HF effect with respect to number of barriers.
Fig.1(d) show no HF effect for high $V_c (=1.9)$}
\label{8.2}
\end{figure}

In evanescent mode of tunneling of waves trough an array of real barrier potentials 
the tunneling time always saturates with the number of barriers. However this is not the case for the 
complex barriers due to the presence of  
absorption or emission in the potential. In the
two channel study the strength of the absorption or emission depends on the potential $V_c$ which
couples  elastic and inelastic channels. We numerically calculate the tunneling time by 
considering  an array of complex barriers with same height, width and with equal spacing.
The effect of $V_c$ on the tunneling time and on absorptivity $(1-R-T)$ is plotted in Fig. \ref{8.2}. We observe
the saturation of tunneling time with respect to the number of barriers when $V_c$ is small and
HF effect disappears for higher value of $V_c$. No saturation of tunneling time with respect to  number of barriers  
occurs when absorption is more in the array of complex barriers. Fig. \ref{8.2} further shows that absorptivity also
saturates with respect to the number of barriers.

%%%%%%%%%%%%%%%%%%%%%%%%%%%%%%%%%%%%%%%%%%%%%%%%%%%%%%%%%%%%%%%%%%%%%%%%%%%%%%%%%%

\pagebreak

\subsection{Dependency of tunneling time on width}
\begin{figure}
\centering 

\hspace{-1.7in} \includegraphics[scale=0.40]{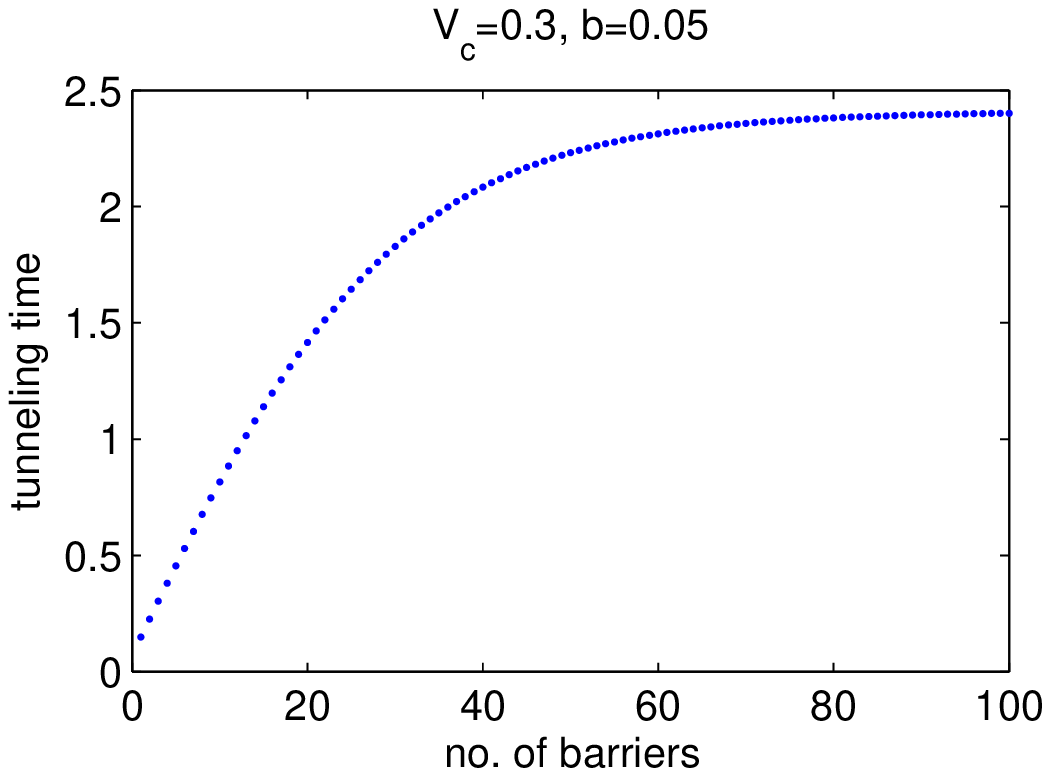} \hspace{-.5in} \includegraphics[scale=.40]{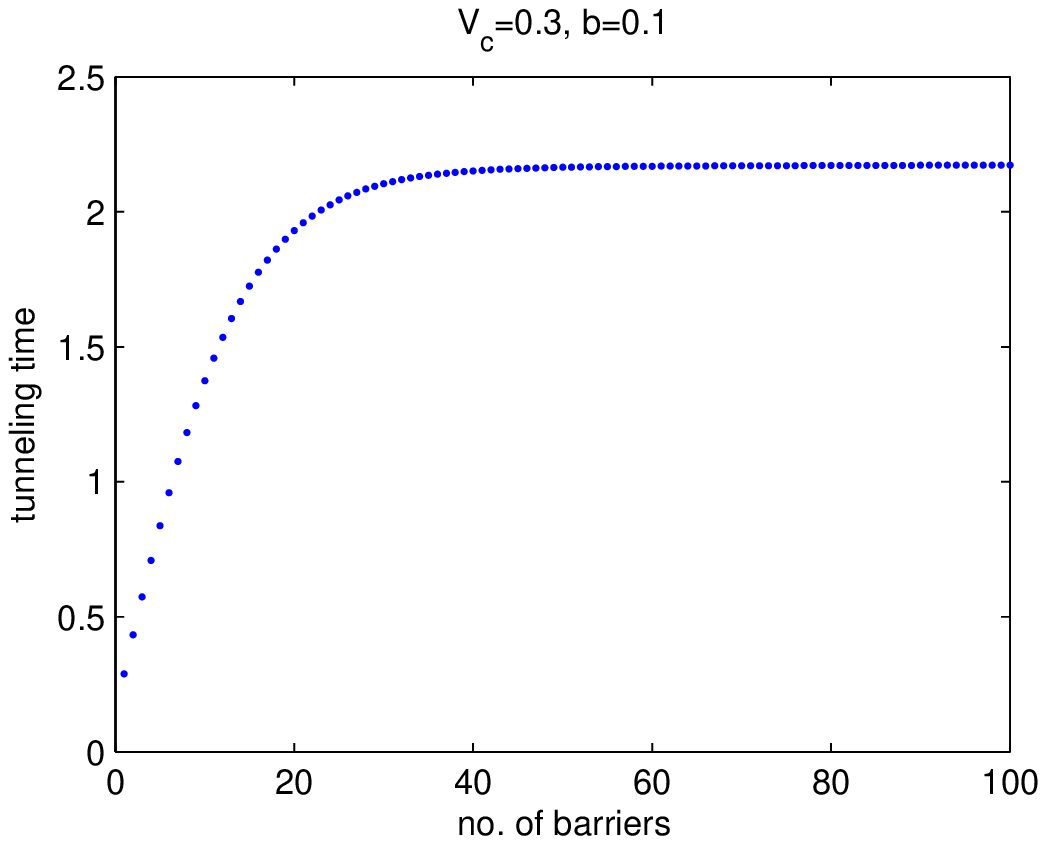} \includegraphics[scale=0.40]{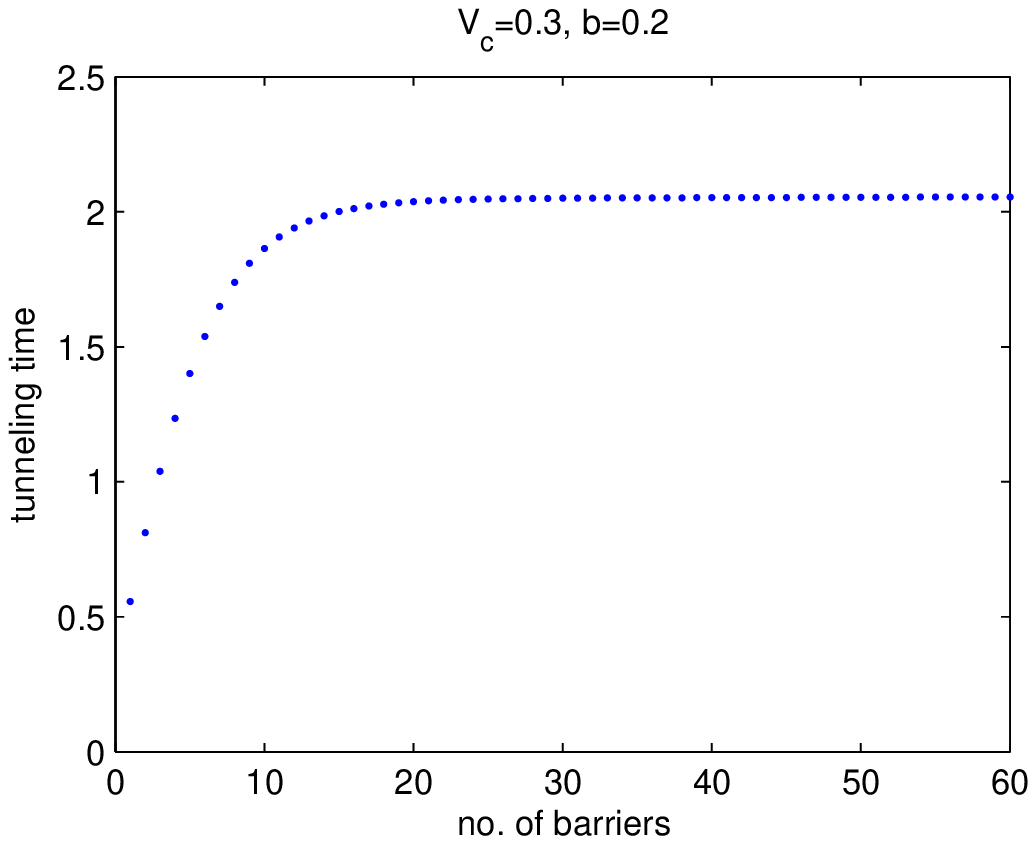} \hspace{-1.8in} 

\hspace{-1.9in} \includegraphics[scale=0.40]{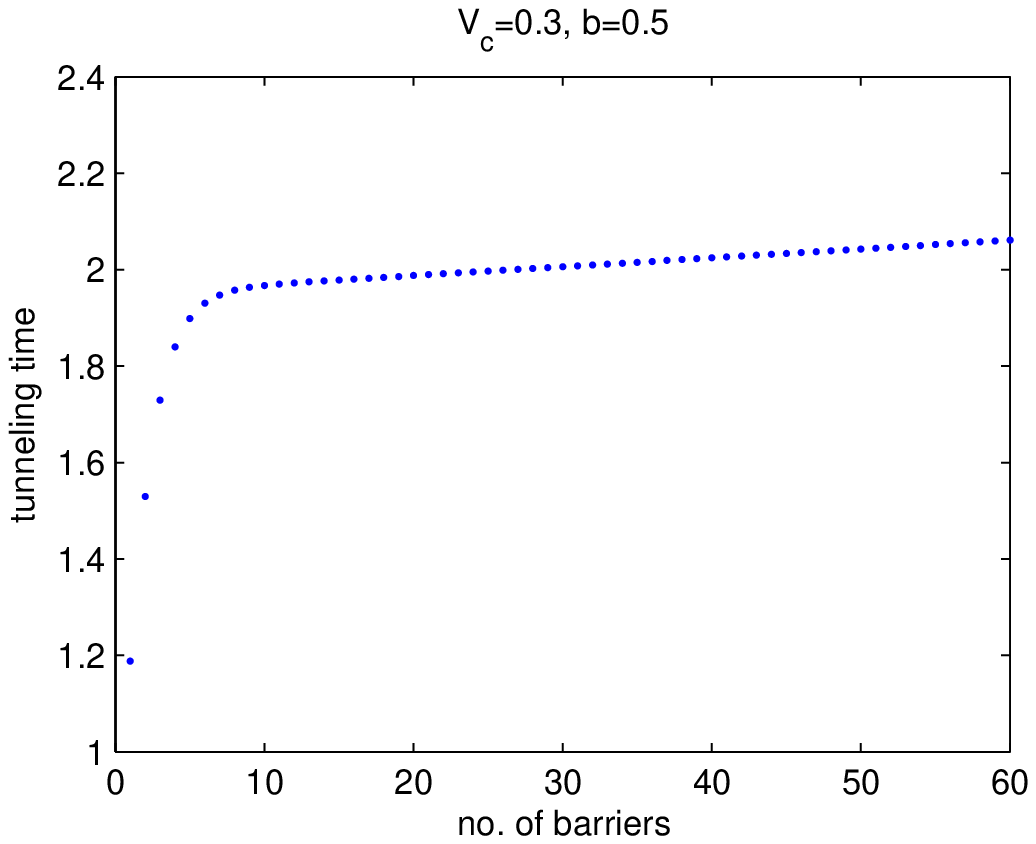} \includegraphics[scale=.40]{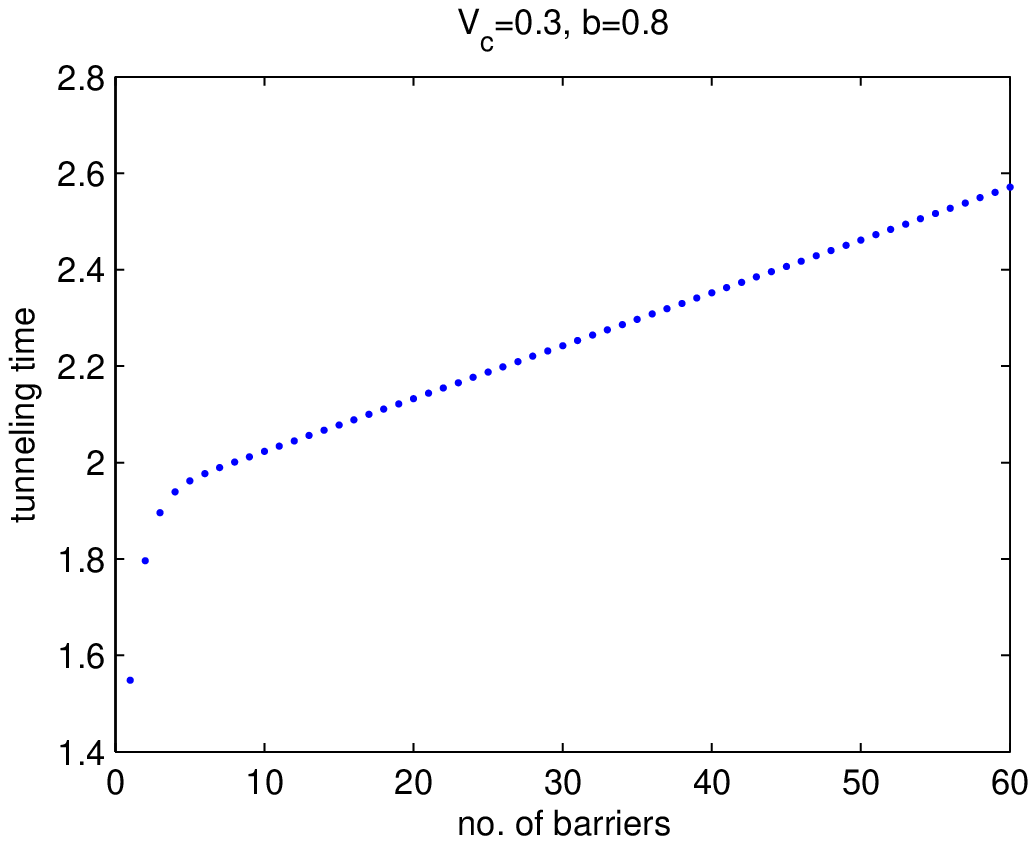} \includegraphics[scale=0.4]{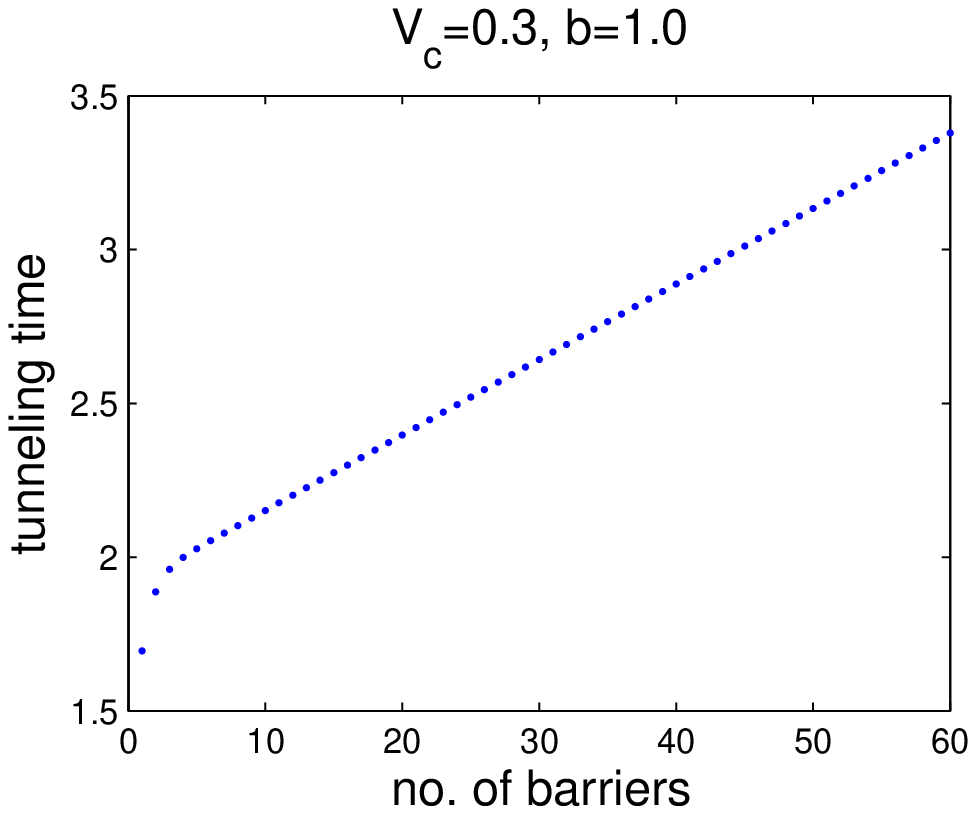} \hspace{-2.2in} 

\caption {\it Tunneling times are plotted with the number of barriers for six different arrays
where the width of the potentials in each array is increasing with a fixed complex 
coupling $V_c=0.3$ and with the same values of the other parameters as in Fig. \ref{8.2}. }
\label{8.3}
\end{figure}

For a single complex barrier with weak coupling tunneling time saturates as  the width of the
barrier increases. In case of the array of the complex barrier with weak coupling we observe HF effect
with respect to number of barrier only when width of the individual barrier is kept between some specific values. This has been demonstrated in Fig. \ref{8.3}. For a fixed weak value of coupling $V_c=0.3$
we plot the tunneling time with respect to the number of the barriers for different values of width 
$b$. We observe HF effect occurs for this particular value of $V_c$ only when width $b$ is
kept between approximately $0.1$ to $0.6$. This behavior of HF effect clearly distinct from that of array of real 
barrier and is demonstrated in Fig. \ref{8.3}.

\begin{figure}
\centering 

 \includegraphics[scale=0.48]{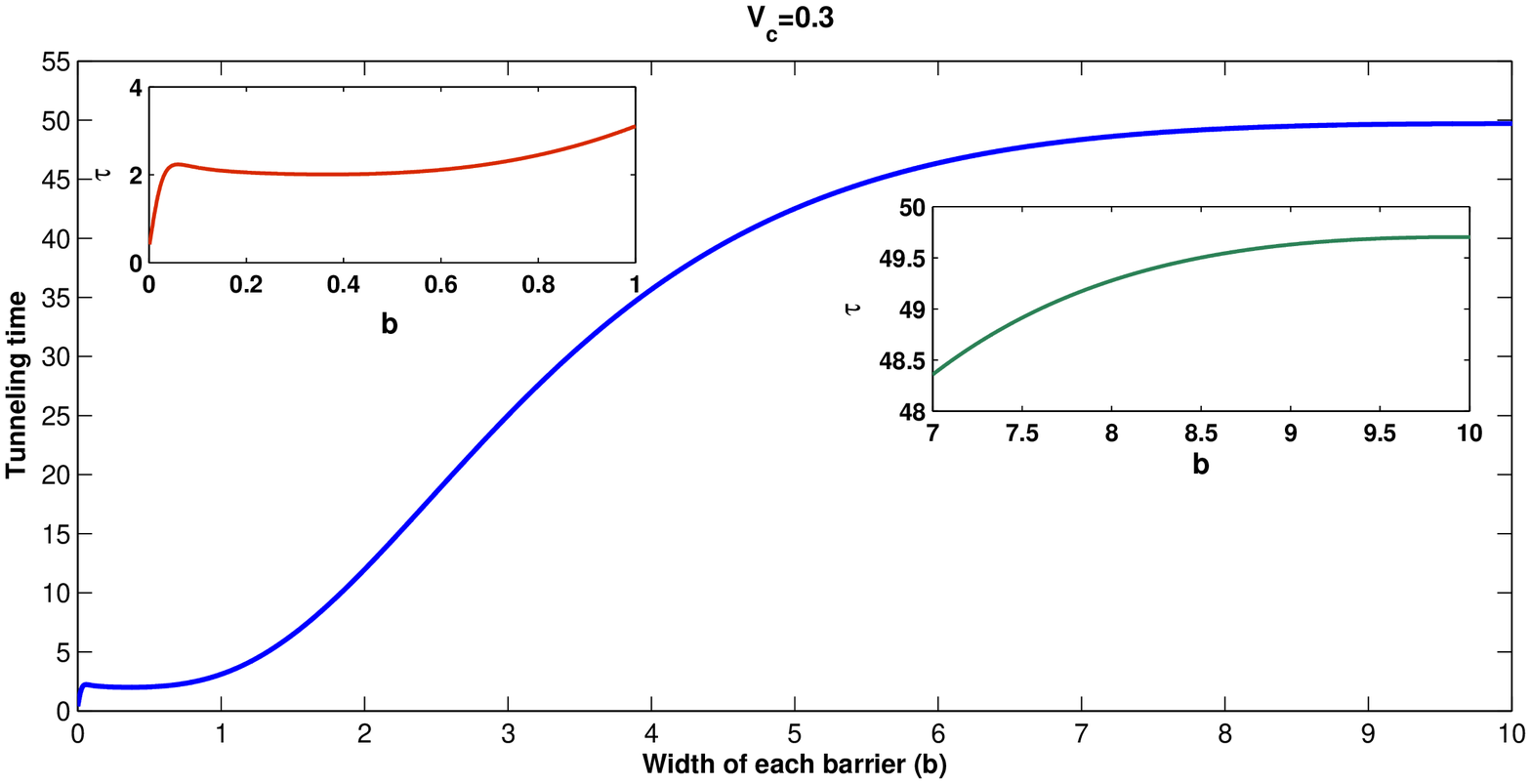} (a)
 
 \vspace{.1in}
 
 \includegraphics[scale=0.48]{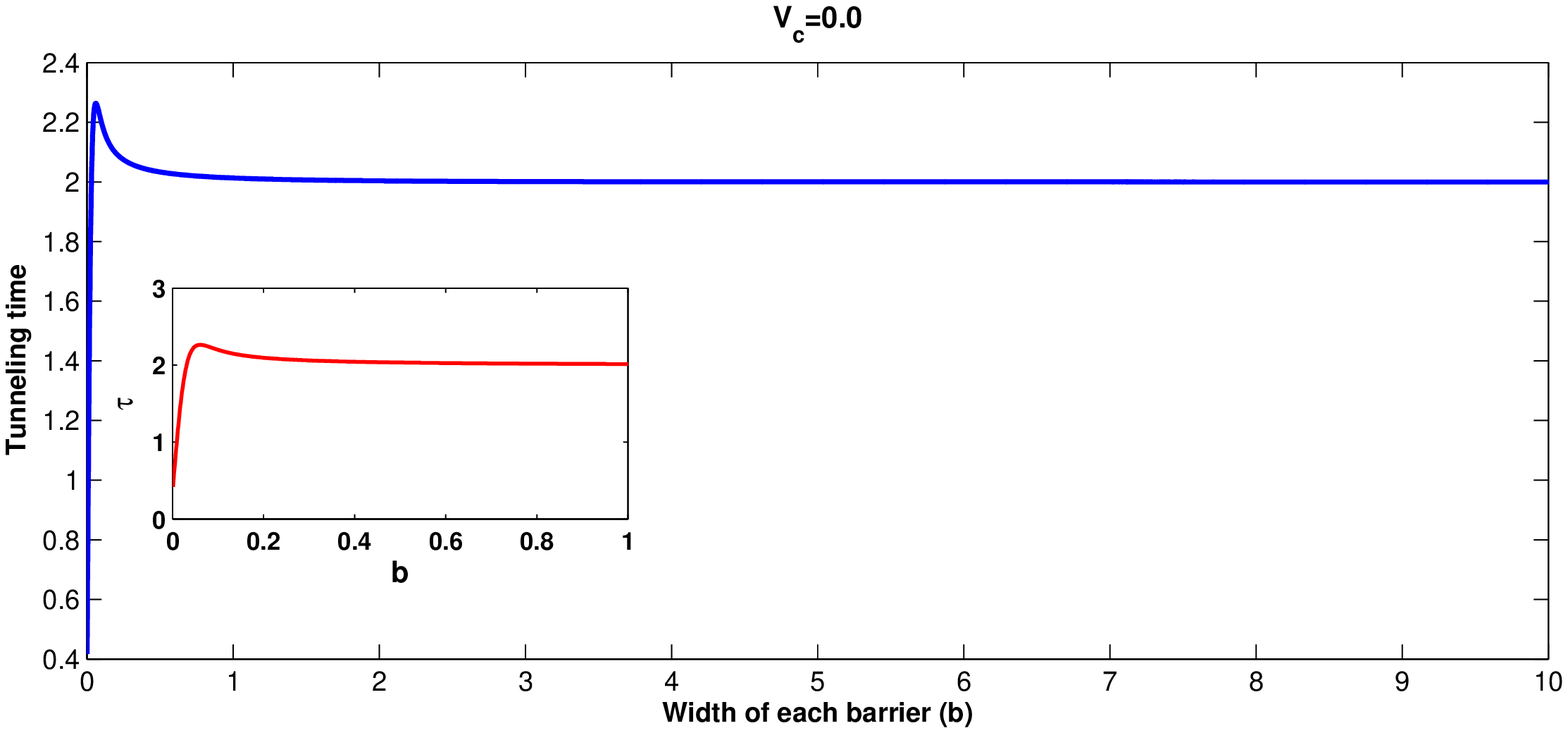} (b)  \\

\caption {\it In (a) and (b) the tunneling times with the width of barriers are plotted for
an array of complex barriers $(V_c=0.3)$ and an array of real barriers $(V_c=0)$ respectively by keeping all other 
parameters same as in Fig. \ref{8.3}. The insets of \ref{8.4}(a) zoom out the behavior of tunneling time for smaller values of width (red line) and lager values of width (green line). 
The same range of width has been enlarged in the inset of \ref{8.4}(b) to capture distinct behavior of tunneling time
for real array. }
\label{8.4}
\end{figure}

In the case of complex barriers saturation of tunneling time occurs for certain ranges of the width of the individual
barriers. We observe HF effect for $0.2<b<0.6$ for $V_c=0.3$ keeping other parameters fixed and tunneling time
starts saturating again when $b>8$
[Fig. \ref{8.4}(a)] \footnote{However for exceptionally 
thick barriers the transmitting wave packet may be distorted.}. 
This result is clearly distinct from the array of real barriers where tunneling time is independent of $b$
except at certain small values [Fig. \ref{8.4}(b)].

%\hspace{-0.5in}\includegraphics[scale=0.46]{contour_n_25.eps} \includegraphics[scale=0.39]{Contour_n_50.eps}

%{\it Fig. 4:}

%%%%%%%%%%%%%%%%%%%%%%%%%%%%%%%%%%%%%%%%%%%%%%%%%%%%%%%%%%%%%%%%%%

\pagebreak

\subsection{Tunneling time resonances} 

\begin{figure}
\centering 

\hspace{-0.4in} \includegraphics[scale=0.55]{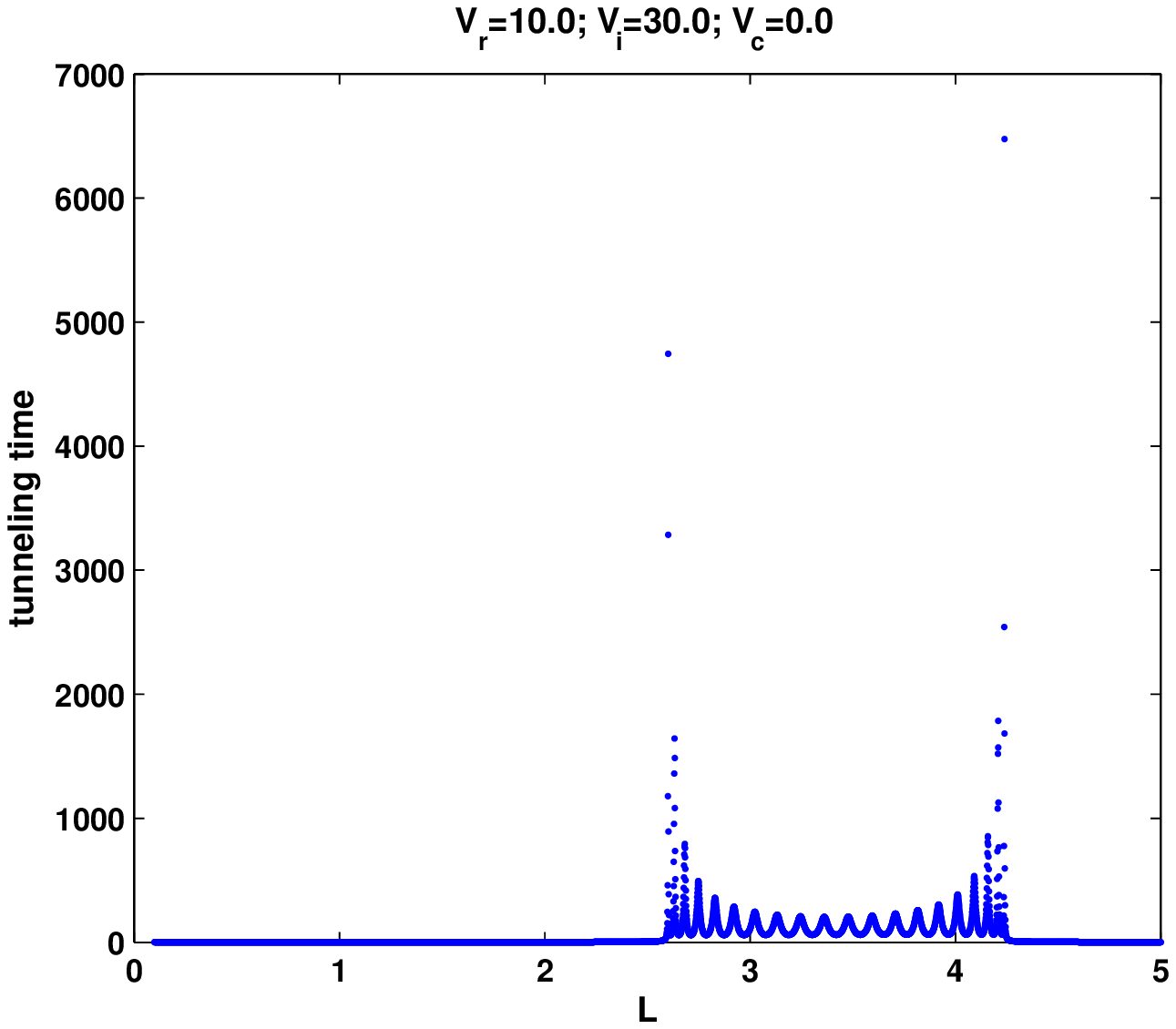} \includegraphics[scale=0.55]{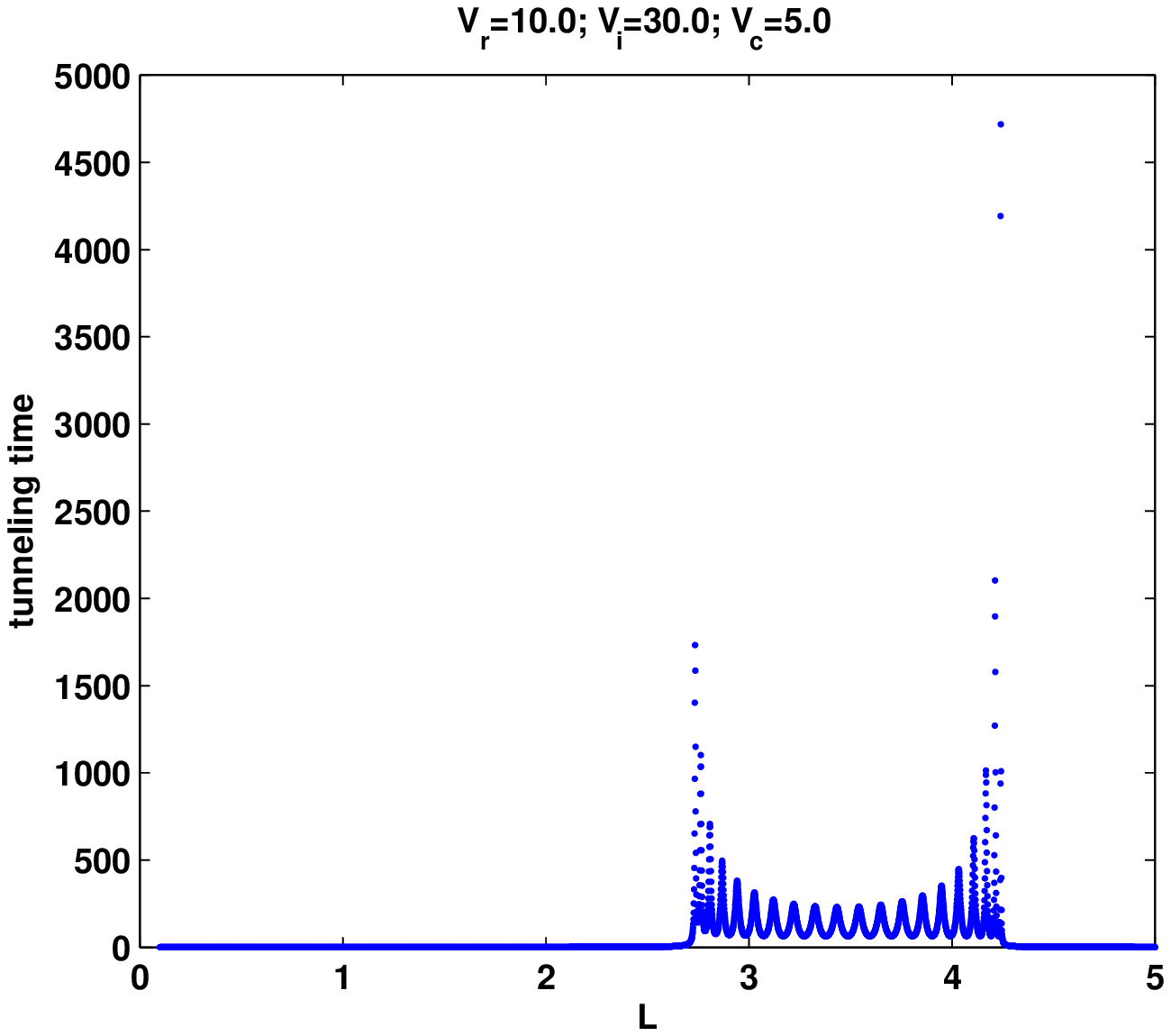} \\

\hspace{-0.4in} \includegraphics[scale=0.55]{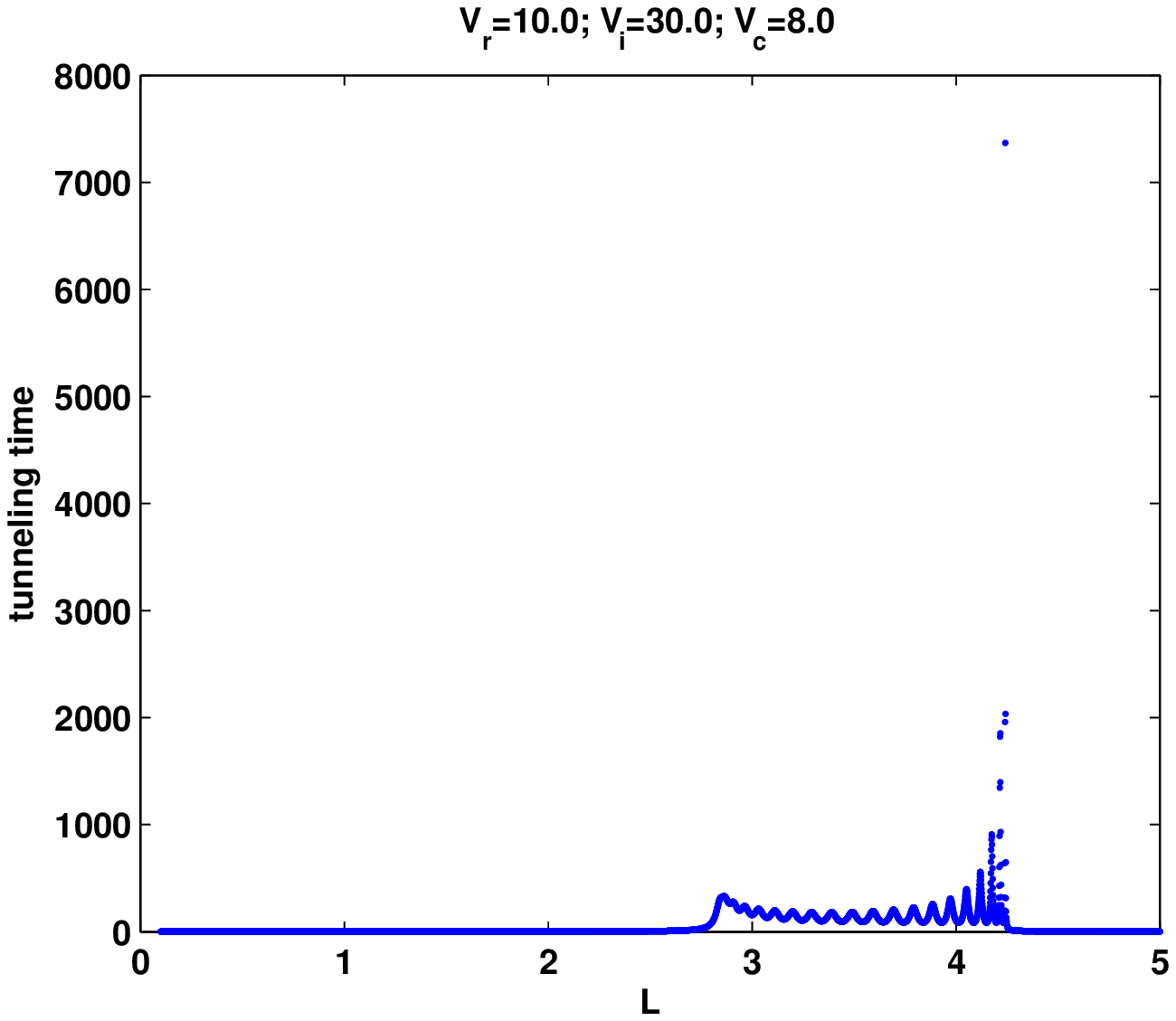} \includegraphics[scale=0.55]{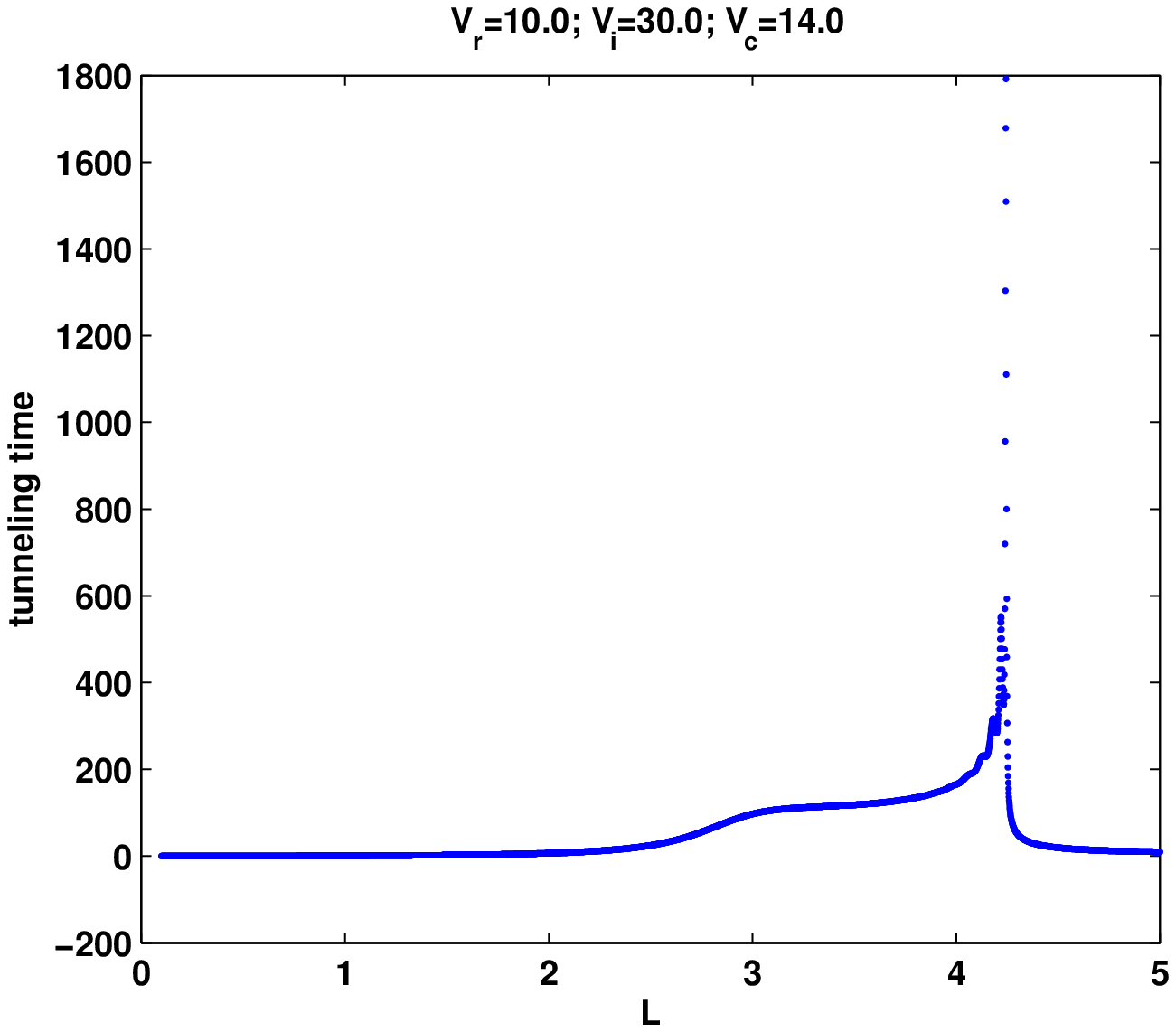}

\caption {\it Tunneling times are shown with consecutive separation of barriers in the array.
Figure (a) is for an array of real barriers and (b,c,d) are for arrays of complex barriers with
different complex coupling $V_c=0, V_c=5, V_c=8$ and $V_c=14$ respectively with the other parameters
same as in Fig. \ref{8.2}. Resonances is regulated as we increase the inelasticity in the system.}
\label{8.5}
\end{figure}

The resonances in tunneling time have been noticed for the case of real barriers \cite{rns1}. 
In case
of one or two real barrier \cite{exp6} it has been reported that for certain values of 
the separation between adjacent barriers the wave packet never emerges
indicating a very high value in tunneling time. In this subsection
we numerically study such resonances in the case of array of complex barriers. In particular we show how 
resonances in tunneling time changes with coupling between elastic and inelastic channels 
and separation between adjacent barriers.  We observe that resonances in tunneling time for
evanescent wave gradually disappear if we increase $V_c$ coupling between the elastic and inelastic channels.
This is explained clearly in Fig. \ref{8.5} by changing $V_c$ in the array of total $20$ real barriers (Fig. \ref{8.5}(a)) and $20$ complex barriers (Fig. \ref{8.5}(b, c, d)).
To see the effect of complex coupling on the resonances we chose a much
opaque barrier by increasing both real and imaginary parts of the barrier height (i.e $V_r, 
V_i$) up to a value so that we can increase the value of $V_c$ without disturbing the 
evanescent mode of tunneling. 
The first resonance disappears when complex coupling $V_c$ raise from $1$ to $14$. Unlike the case of real barriers 
the resonances in the case of array of complex barrier is regulated by increasing inelasticity.  \\

\pagebreak

\subsection{HF effect in randomly arranged arrays} 

\begin{figure}
\centering

\hspace{-0.3in}\includegraphics[scale=0.52]{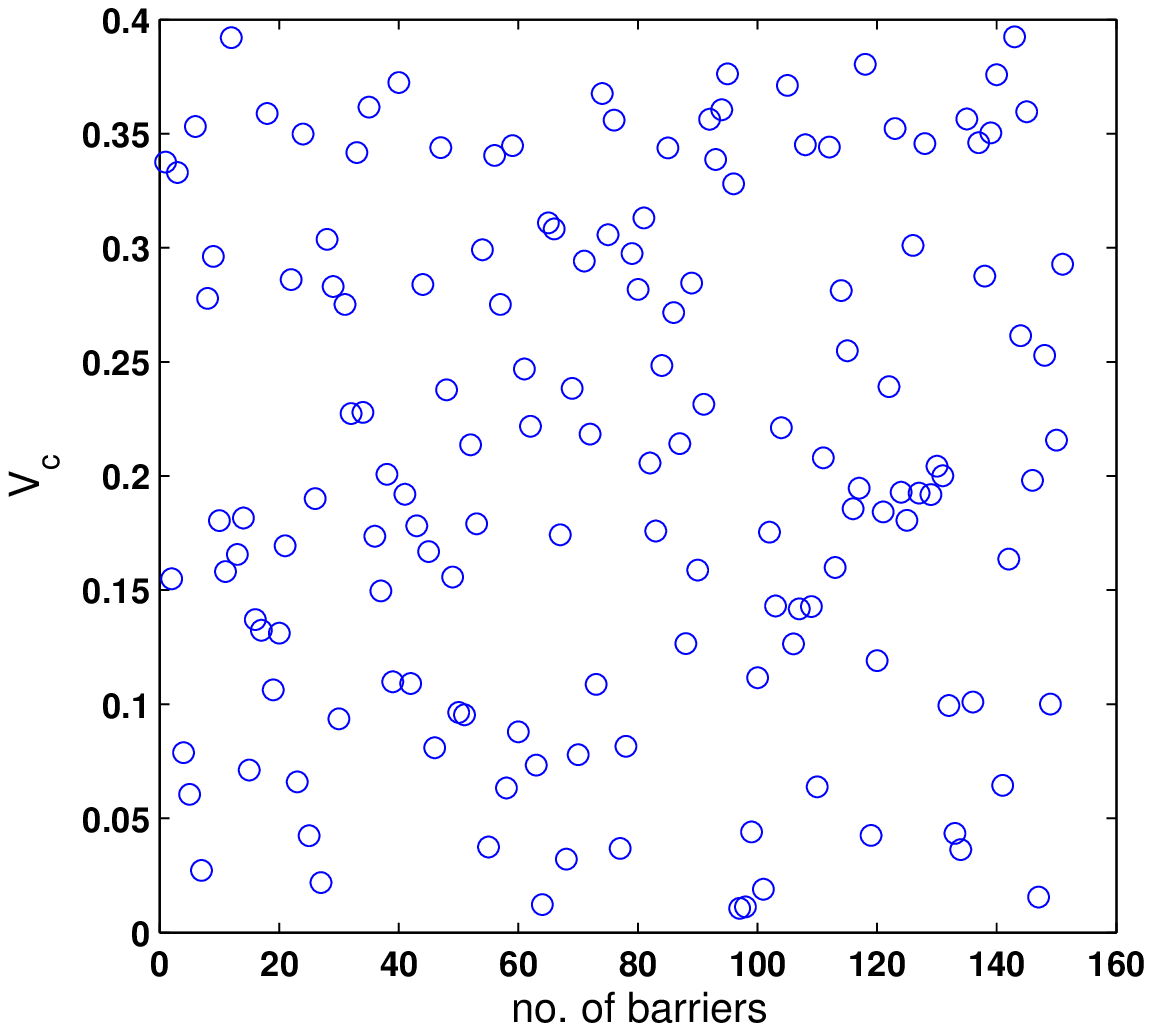} (a) \includegraphics[scale=0.5]{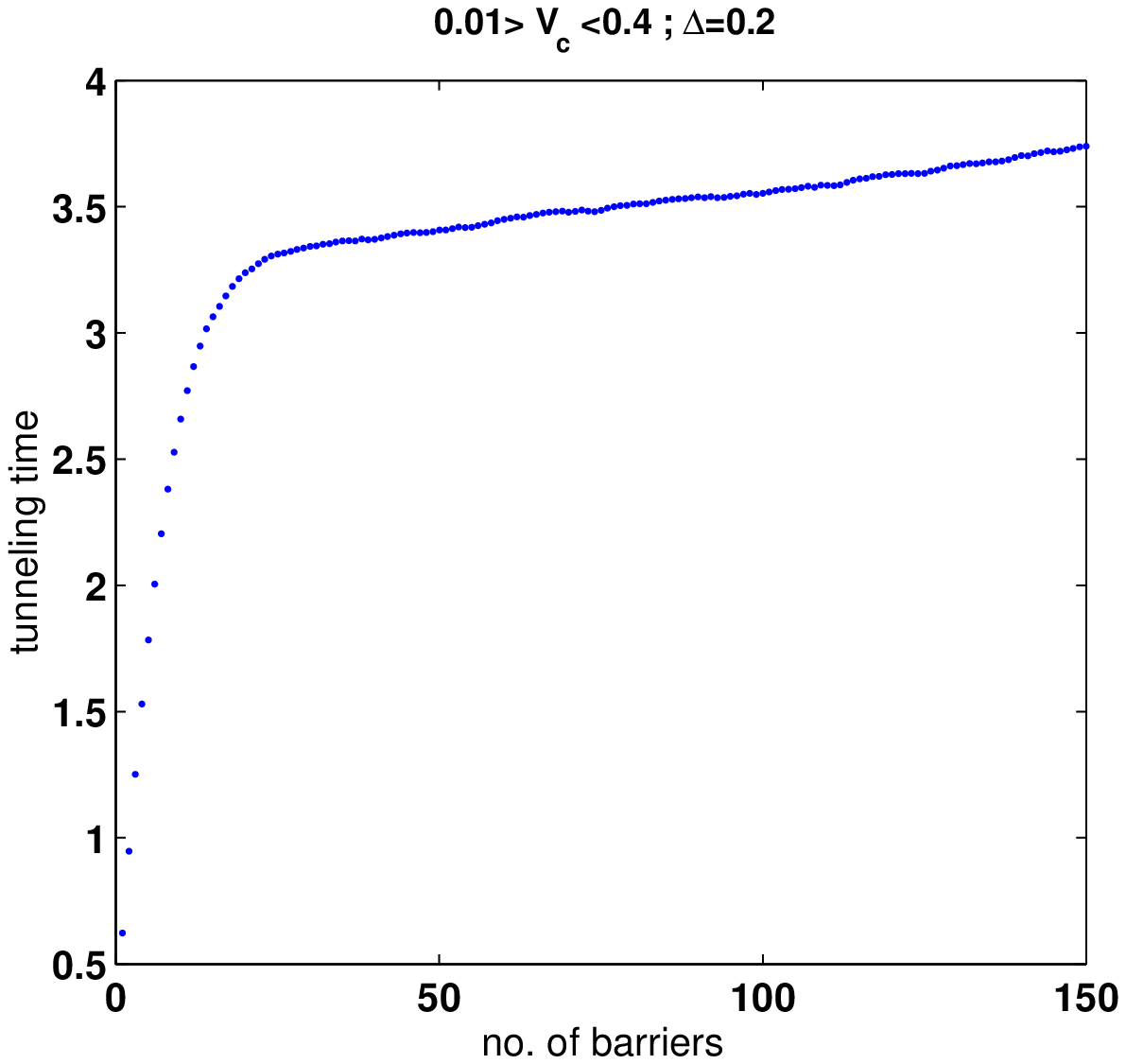} (b)

\hspace{-0.3in}\includegraphics[scale=0.52]{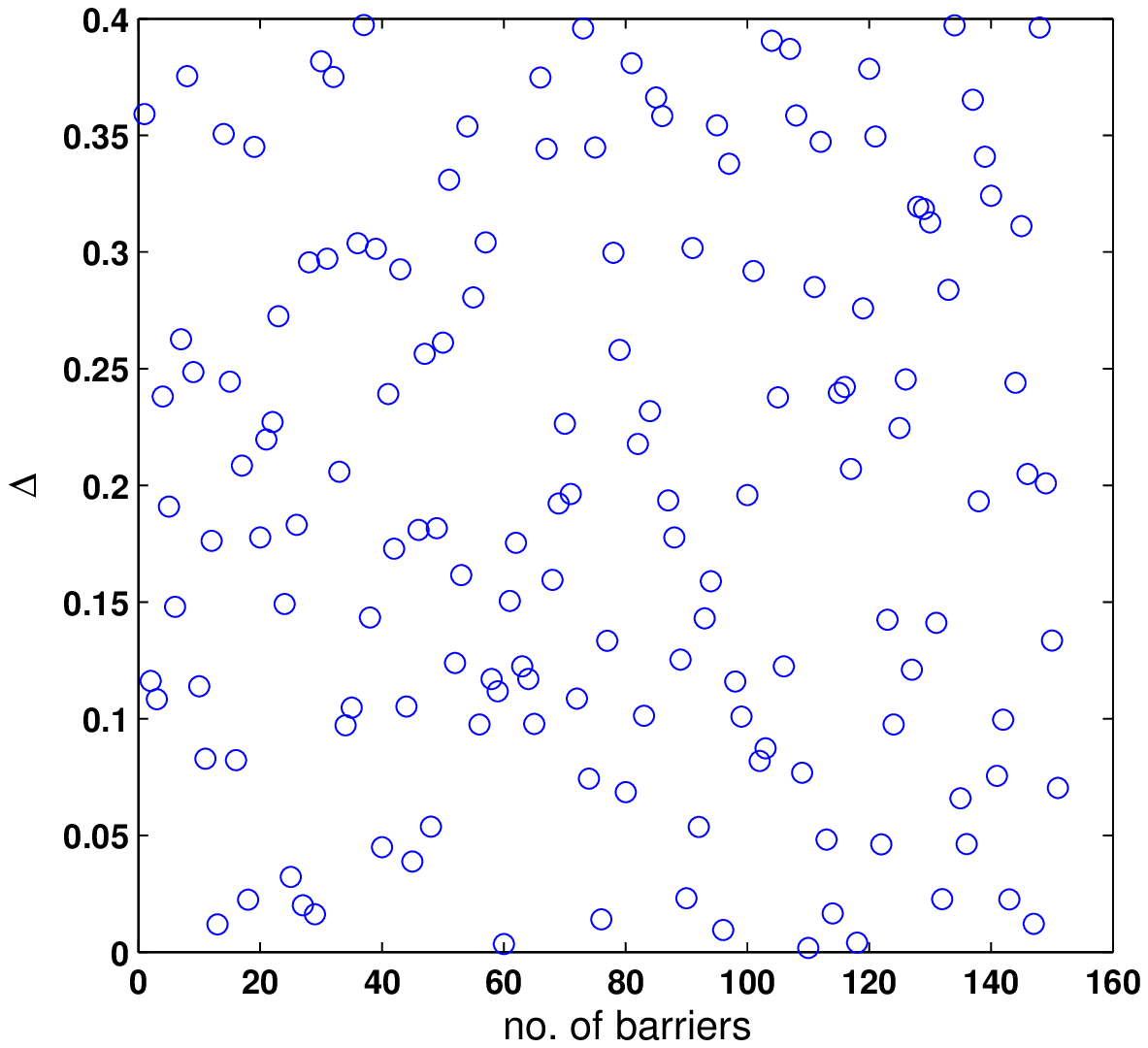} (c) \includegraphics[scale=0.52]{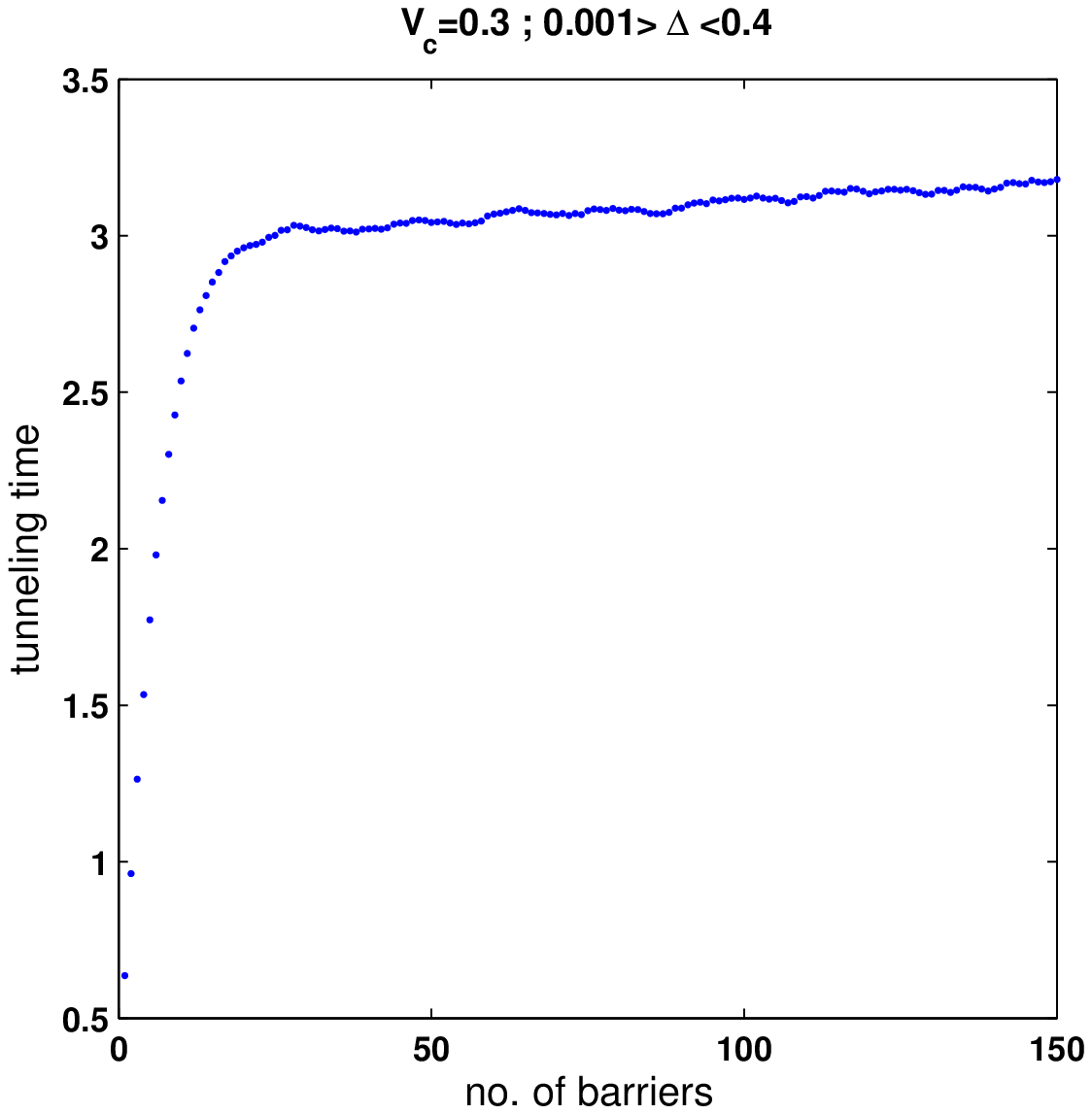} (d) \\

\caption {\it The random distributions (between a range of 0.01 to 0.4) of coupling $V_c$ and 
excitation energy $\Delta $ of each barriers in two different arrays have been shown in (a) and (c). The 
corresponding tunneling times for these two arrays are shown in the figures (b) (for fixed $\Delta =0.02 $ ) and (d) 
(for fixed $V_c =0.3 $ ) respectively. }
\label{8.6}
\end{figure}

In this section we would like to point out HF effect is not an artifact of regularity in terms of
coupling $V_c$ and/ or $\Delta$. In a realistic system the inelasticity may be different in different barriers.
Keeping this in view we chose the coupling $V_c$ randomly for different complex barriers in the array and
study tunneling time by considering arbitrary excited state energies $(\Delta)$.
We show HF effect exists even for such systems and are demonstrated in Fig. \ref{8.6} (a, b). Similarly
we consider another array of barriers in which individual barriers correspond to different inelastic channel with random values of $\Delta$. We show the saturation in tunneling time in such a array
of barrier for the propagation of wave packet in the evanescent mode. This is graphically 
represented in Fig. \ref{8.6} (c, d) with keeping the other parameters same as Fig. \ref{8.2}.

\pagebreak

\subsection{Emissive inelastic channel} 
\begin{figure}
\centering

\includegraphics[scale=0.50]{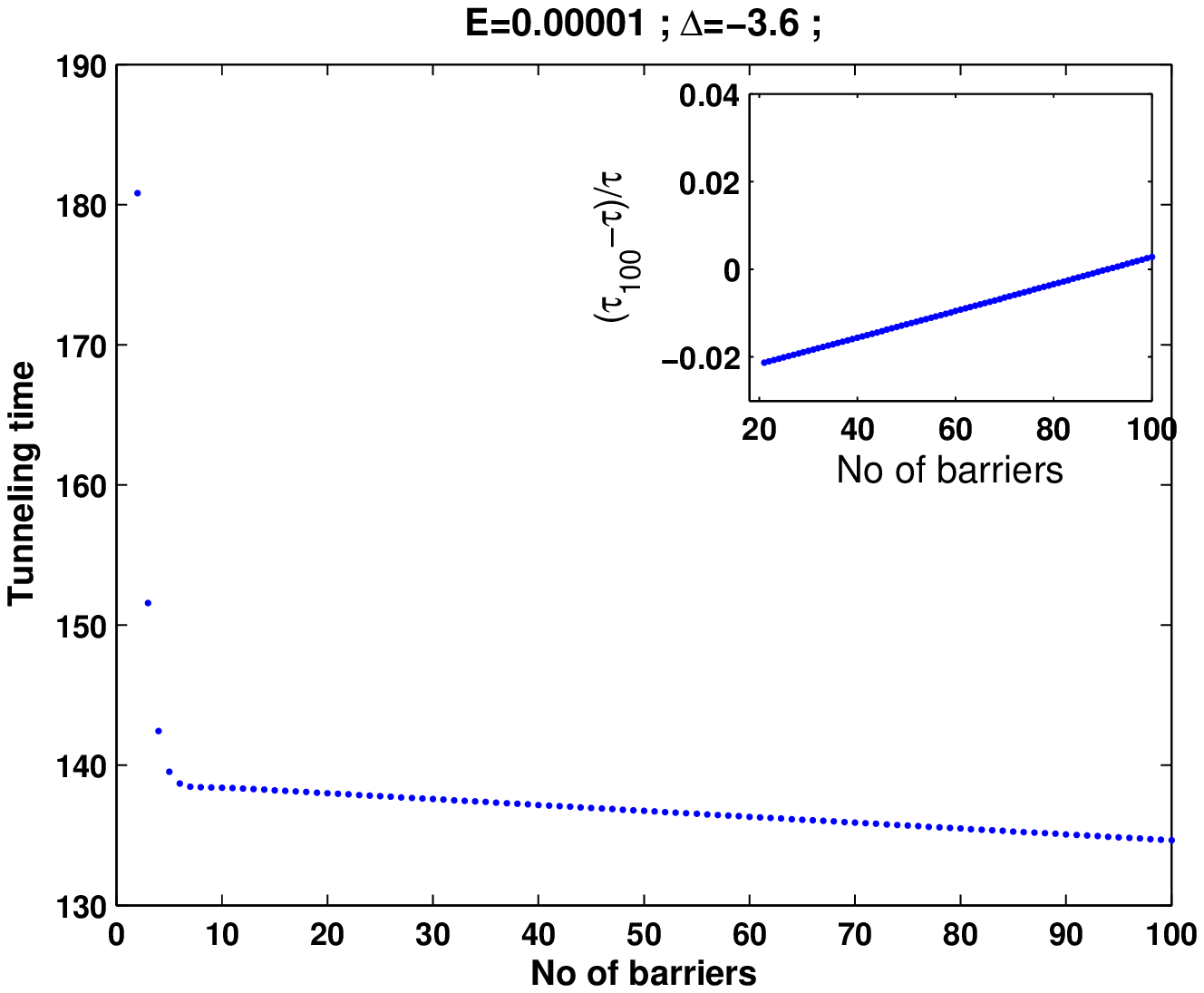} \includegraphics[scale=0.50]{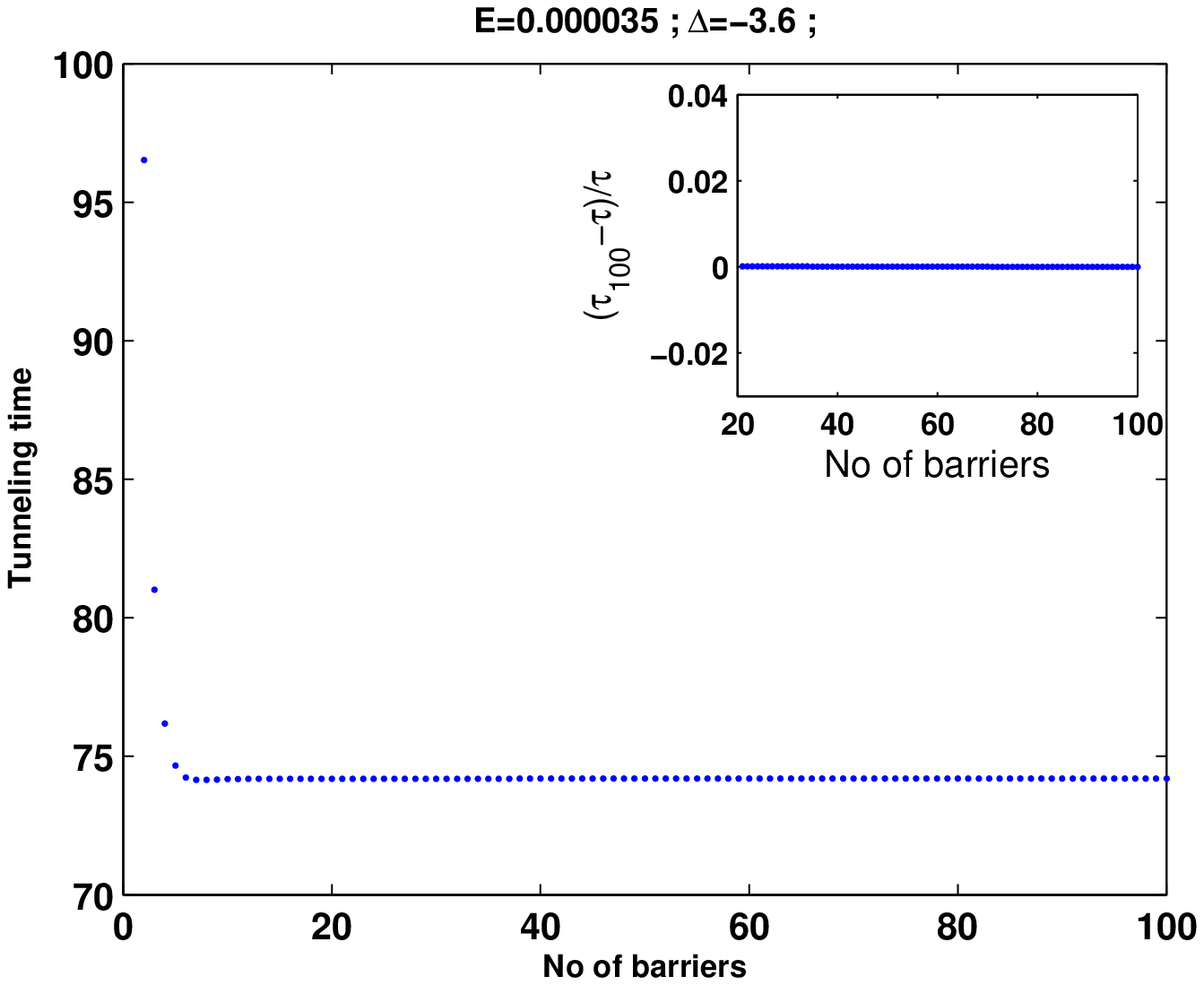}

\includegraphics[scale=0.50]{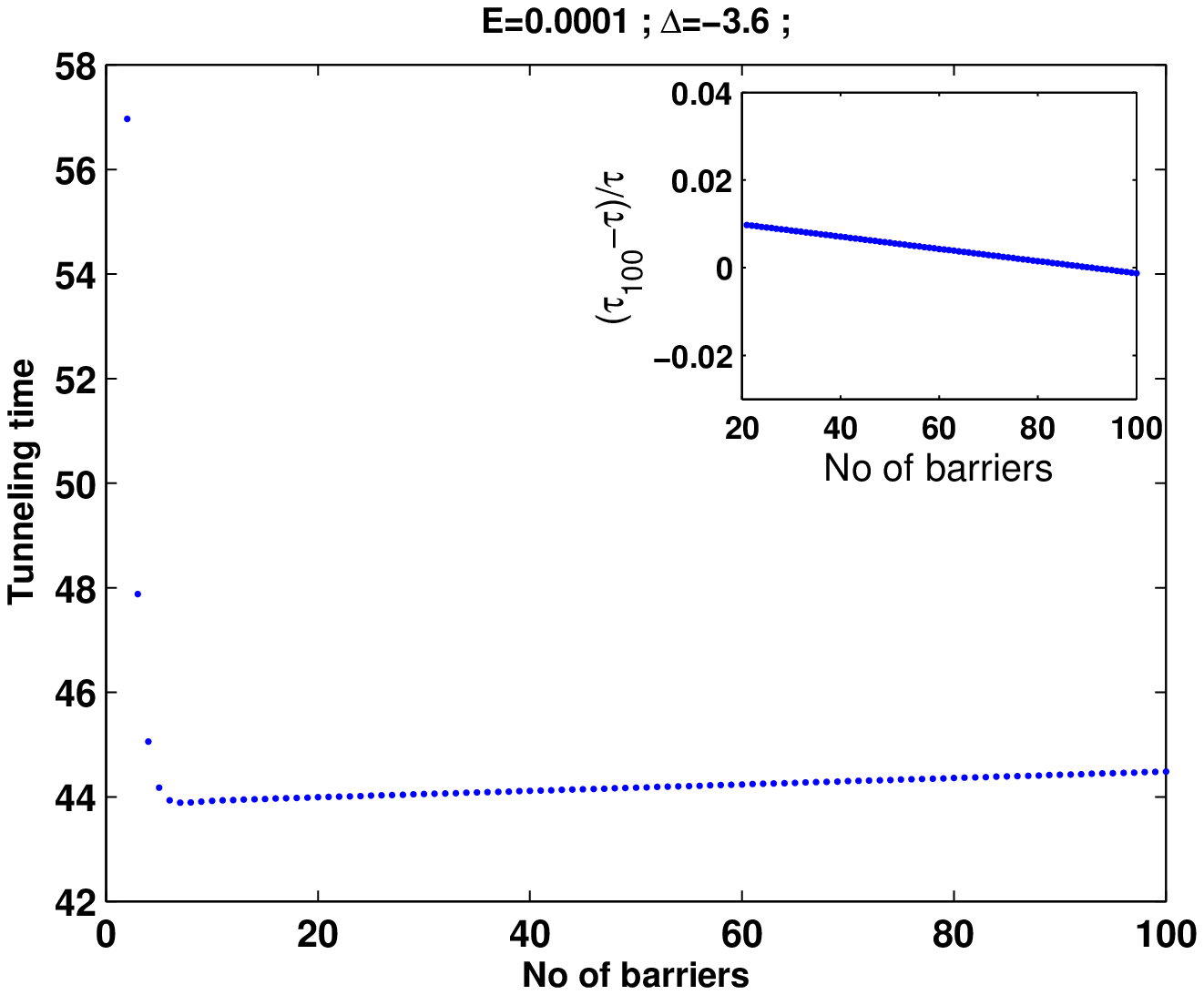} \includegraphics[scale=0.50]{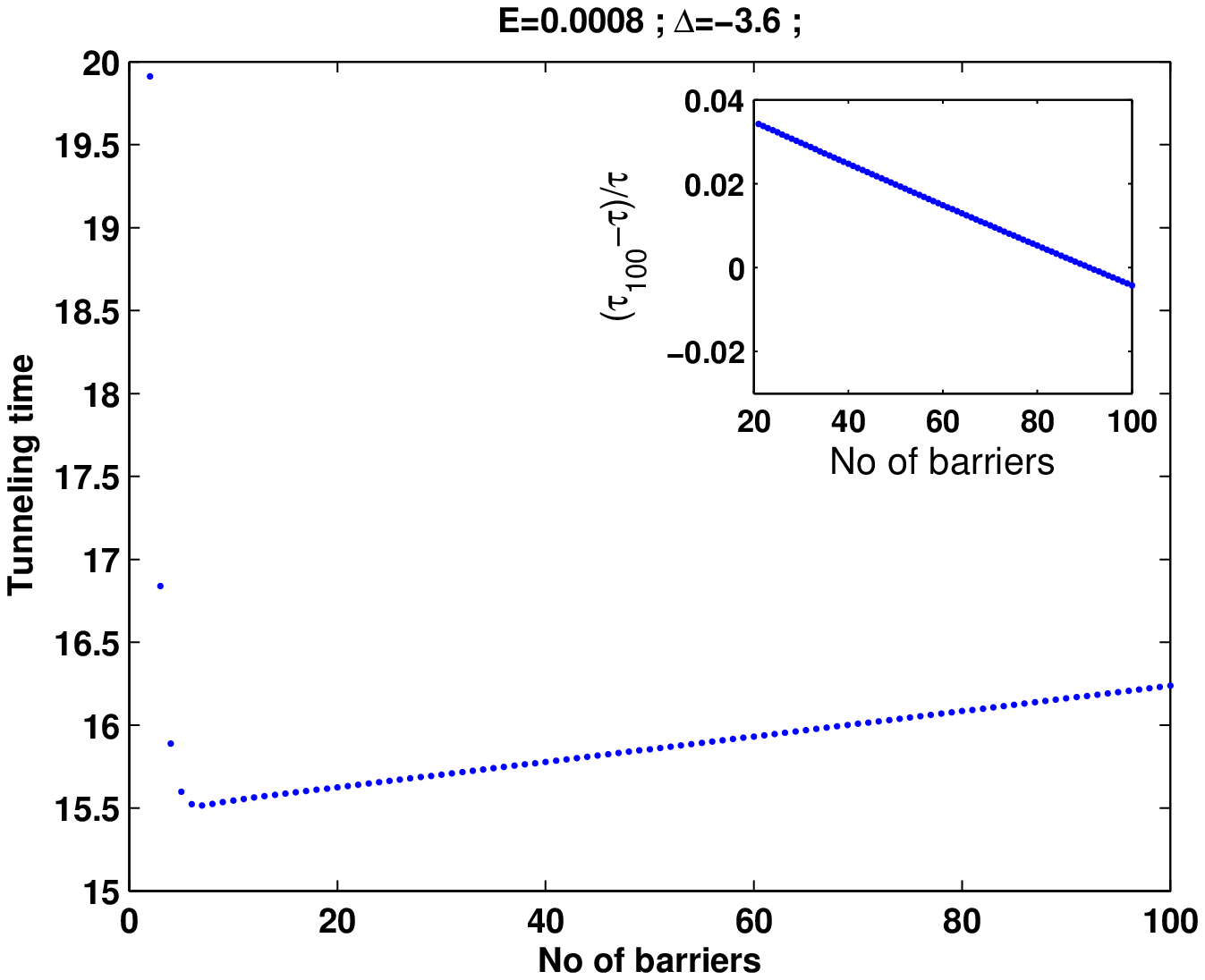}

%\hspace{-0.4in} \includegraphics[scale=0.40]{spem_pll01.eps} \includegraphics[scale=0.40]{spem_p1.eps}\includegraphics[scale=0.40]{spem_3p.eps} 

\caption {\it  Tunneling times with the number of barriers in an array 
for fixed $\Delta=-3.6$ (with $ V_r=10, V_c=14$ and other values are same as Fig. \ref{8.2}) are shown when
the incident particle energy is considered very small, from a value of $E=0.00001$ to $E=.0008$. }
\label{8.7}
\end{figure}

In the previous subsections we consider absorptive ($\Delta$ is positive) inelastic channel to demonstrate HF effect. Now we concentrate emissive inelastic channel by considering target in an excited state $\Delta$. 
The Schroedinger equation in this inelastic channel 
is then written by replacing $\Delta$ by $-\Delta$ in Eq. (\ref{inel}).
Therefore an emission is being associated with the interaction of waves and the potential.
For $E<<\Delta$ both elastic and inelastic processes occur in presence of such a target. We observe HF effect
with respect to number of barriers at particular very low energy of the incident wave packet (see Fig. \ref{8.7}). 
In the inset of each graphs of Fig. \ref{8.7} we have plotted the relative tunneling time 
$(\frac{\tau_{100}-\tau}{\tau})$ 
with respect to an array of $100$ barriers. Fig. \ref{8.7}(b) shows the HF effect at around 
$E=.00035$ for $\Delta=3.6$. Occurrence of HF effect at particular low incident energies is an interesting result in 
the case of array of complex barriers and need further investigation.

\section{Conclusions}
We have studied the HF effect for an array of complex barrier potentials to unfold various new features associated with
this interesting effect by using stationary phase method. We have constructed the total scattering transfer matrix by multiplying the
transfer matrices for the individual barriers to calculate the tunneling time for a wave packet through such an array of barriers.
The tunneling time saturates with respect to the number of barriers depending on the different parametric values in the system.
Saturation crucially depends on the coupling potential $V_c$ which couples elastic and inelastic channels of propagation when other parameters 
are held fixed. We have observed HF effect only for low absorption (i.e. small $V_c$) in the system. 
Saturation in tunneling time with
respect to number of barriers are observed for certain ranges of the width of barrier unlike the situation for the 
real array where saturation is always achieved beyond a certain value of the width. In case of real array for certain 
values of the width of the barrier and separation of adjacent barrier the tunneling time is extremely 
high and the wave packet never emerges from such barriers. We have shown such a resonance behavior in tunneling time 
is regulated in case of array of complex barriers. For the sake of realistic systems
we further have studied the HF effect for the array of barriers with random value of coupling with arbitrary
value of excited energy $(\Delta)$ and with fixed inelasticity with random $(\Delta)$. In both cases we have shown the saturation of 
tunneling time with respect to number of barriers. Finally we have observed HF effect even for the case of emissive inelastic channel. 
Surprisingly saturation of tunneling time occurs there at some particular very low incident energy. This needs further investigations.

\vspace{.25in}

{\it Acknowledgment:} BPM acknowledges the financial support from the Department 
of Science \& Technology (DST), Govt. of India, under SERC project sanction grant No. 
SR/S2/HEP-0009/2012 and MH is thankful to Dr. S. K. Shivakumar, Director, ISAC for his 
support to carry out this research work. AG acknowledges the Council of Scientific \& 
Industrial Research (CSIR), India for Senior Research Fellowship.

\end{document}